\documentclass[modern]{aastex62}

\newcommand{\acronym}[1]{{\small{#1}}}
\newcommand{\CRLB}{\acronym{CRLB}}
\newcommand{\FF}{\texttt{FAST FORWARD}}
\newcommand{\package}[1]{\textsl{#1}}

\shorttitle{information in stellar streams}
\shortauthors{bonaca \& hogg}

\usepackage{amsmath}

\begin{document}\sloppy\sloppypar\raggedbottom\frenchspacing 

\title{The information content in cold stellar streams}

\correspondingauthor{Ana Bonaca}
\email{ana.bonaca@cfa.harvard.edu}

\author[0000-0002-7846-9787]{Ana Bonaca}
\affil{Harvard--Smithsonian Center for Astrophysics}

\author[0000-0003-2866-9403]{David W. Hogg}
\affiliation{Center for Cosmology and Particle Physics,
Department of Physics,
New York University}
\affiliation{Center for Data Science,
New York University}
\affiliation{Max-Planck-Institut f\"ur Astronomie, Heidelberg}
\affiliation{Flatiron Institute, Simons Foundation}

\begin{abstract}\noindent 
Cold stellar streams---produced by tidal disruptions of clusters---are long-lived, coherent dynamical features in the halo of the Milky Way.
They have delivered precise information about the gravitational potential, including constraints on the shape of the dark-matter halo.
Because of their different ages and different positions in phase space, different streams tell us different things about the Galaxy.
Here we employ a Cram\'er--Rao (\CRLB) or Fisher-matrix approach to understand the quantitative information content in (toy versions of) eleven known streams: ATLAS, GD-1, Hermus, Kwando, Orinoco, PS1A, PS1C, PS1D, PS1E, Sangarius and Triangulum.
This approach depends on a generative model, which we have developed previously, and which permits calculation of derivatives of predicted stream properties with respect to Galaxy and stream parameters.
We find that in simple analytic models of the Milky Way, streams on eccentric orbits contain the most information about the halo shape.
For each stream, there are near-degeneracies between dark-matter-halo properties and parameters of the bulge, the disk, and the stream progenitor itself, but simultaneous fitting of multiple streams will constrain all parameters at the percent level.
At this precision, simulated dark matter halos deviate from simple analytic parametrizations, so we add an expansion of basis functions to give the gravitational potential more freedom.
As freedom increases, the information about the halo reduces overall, and it becomes more localized to the current position of the stream.
In the limit of high model freedom, a stellar stream appears to measure the local acceleration at its current position; this motivates thinking about future non-parametric approaches.
The \CRLB\ formalism also permits us to assess the value of future measurements of stellar velocities, distances, and proper motions.
We show that kinematic measurements of stream stars are essential for producing competitive constraints on the distribution of dark matter, which bodes well for stream studies in the age of \textsl{Gaia}.
\end{abstract}


\section{Introduction} \label{sec:intro}
Immediately following the discovery of the first accreted structures in the Milky Way's halo \citep{ibata1994, ti1998}, \citet{johnston1999} recognized that these tidal tails can inform about the properties of the halo itself, concluding \emph{``you can judge a galaxy by its tail''}.
They simulated tidal disruption of a Sagittarius-like satellite in different models of the Milky Way, and showed that with precise observations of the stream's positions and kinematics, the mass and shape of the halo can be constrained with a percent precision.
Despite the urgency to understand the distribution of matter in the Galaxy, in particular the dark matter-dominated halo, it took several years for the comprehensive models of the Sagittarius system to emerge \citep{helmi2004, johnston2005, lm10}.
Gathering the required observational data was the main cause of the delay -- streams are diffuse and faint, so it is challenging to isolate stream members from the overwhelming Milky Way field.

To this day, a total of four streams have been used to measure the Milky Way's gravitational potential: tidal tails of the Sagittarius dwarf galaxy \citep{lm10,gibbons2014,dl2017}, Orphan stream \citep{newberg2010}, GD-1 stream \citep{koposov2010, bowden2015} and tails of Palomar~5 globular cluster \citep{kupper2015}.
A consensus among the nearby streams is that the inner halo is nearly spherical \citep{bovy2016}, however there is no global solution that simultaneously reproduces properties of all streams \citep{pearson2015}.
The inference of halo properties in its outskirts is particularly sensitive to the adopted parametric form; for example, there are equally viable models of the Sagittarius stream in a triaxial halo \citep{lm10}, a halo interacting with the Large Magellanic Cloud that is oblate in the inner part and only mildly triaxial in the outer part \citep{vch2013} and a spherical halo with non-monotonically decreasing density profile \citep{ibata2013}. 

Stream studies have so far adopted simple, analytic forms to describe the dark matter halo because the current data can reliably constrain only a limited number of free parameters.
However, N-body simulations of galaxy formation predict more complex halos \citep[e.g.,][]{diemand2007,springel2008,wetzel2016}, 
so in \citet{bonaca2014} we tested how accurately the streams that were simulated in a cosmological dark matter halo can be modeled in static, analytic potentials.
We found that on average, a population of streams recovers the true halo mass as well as allowed by the analytic model (which itself can be biased a few tens of percent), but that individual streams can be extremely biased, thus posing the question: what are the stellar streams actually measuring?

To quantitatively address this question, we here build a framework for measuring the information content in stellar streams regarding the gravitational potential.
Our goals are two-fold: (1) given a simple parametric model of the Galaxy, we want to know what kind of data we need to optimally constrain it, and more formally (2) what aspects of a non-parametric gravitational potential individual streams constrain.
We use the formalism of Cram\'er--Rao lower bounds and describe how we adopted it to stellar streams in Section~\ref{sec:method}.
We studied a sample of Milky Way-like streams (described in Appendix~\ref{sec:streams}), assuming current or near-future observational data sets (\S\,\ref{sec:datasets}).
In Section~\ref{sec:results} we present how well streams constrain parameters of a simple potential, both individually (\S\S\,\ref{sec:res_ind},~\ref{sec:res_comp}) and jointly (\S\,\ref{sec:res_joint}).
Exploring these constraints in a bit more detail, we show that data of higher dimensionality provide superior constraints to data of higher precision but lower dimensionality (\S\,\ref{sec:forecast}).
Finally, relaxing our model for the gravitational potential, we show that streams best constrain the radial acceleration at their present location (\S\S\,\ref{sec:interpretation},~\ref{sec:bfe}).
In our final section~\ref{sec:discussion}, we discuss the implications of these results for future studies endeavoring to constrain the Milky Way potential using stellar streams.

\section{Methods}
\label{sec:method}

\subsection{Information content in stellar streams}
Numerous methods have been developed to estimate properties of a dark matter halo by modeling observations of stellar streams \citep[e.g.,][]{varghese2011,sanders2013,bonaca2014,bovy2014,apw2014}.
Traditionally, these describe the Galaxy as an analytic model with a handful of free parameters.
Therefore, we define the information content in stellar streams as the best-case uncertainties on the model parameters achievable using the observational data at hand.
Formally, the lower bound on the variance of an unbiased frequentist estimator of a deterministic parameter is given by the Cram\' er--Rao lower bound \citep[\CRLB,][]{Cramer1946, Rao1945}.

For some data set $\vec{y}$ (e.g., a vector of position and velocity measurements of stars along a stream), the associated covariance matrix is $C_y$.
Given the model parameters $\vec{x}$ (e.g., a vector with the mass and scale radius of a dark matter halo), the Cram\' er--Rao bound (under the assumptions that the noise is Gaussian and the model predictions are continuous) becomes the covariance matrix for the model parameters $C_x$, which is a local linear transformation of the covariance matrix for the data $C_y$.
That is, the \CRLB\ for vector $\vec{x}$ is a covariance matrix or variance tensor, which itself is the inverse of a Fisher information matrix \citep{fisher}:
\begin{equation}
C_x^{-1} = \left(\frac{d\vec{y}}{d\vec{x}}\right)^{T} C_y^{-1} \left(\frac{d\vec{y}}{d\vec{x}}\right) + V_x^{-1}
\label{eq:crlb}
\end{equation}
where the derivative object is a rectangular matrix, and $V_x$ is a covariance matrix representing any prior knowledge of model parameters (which, strictly, ought to be likelihood information or a Fisher matrix from prior data).

Although strictly the \CRLB\ is a constraint on frequentist estimators, it also has an interpretation in Bayesian inference.
Heuristically, it gives the variance of the posterior probability distribution function (pdf) when the data are very informative, or in the limit of very good data.
It also delivers the variance of the posterior pdf precisely in the magic situation in which the model is linear (derivatives $d\vec{y}/d\vec{x}$ don't vary with parameters $x$) and both the noise and the prior pdf are Gaussian (and that prior variance is included as $V_x$ in the calculation), but that situation does not strictly apply here.
Nonetheless, it is a first-order approximation to the uncertainties that would be obtained in a sensible Bayesian inference with good data, which is relevant to this work.

Through the rest of this section, we describe individual terms of Equation~\ref{eq:crlb} in the context of stellar streams:
First we present the model of the Galaxy and the existing constraints on its components (\S\ref{sec:model}).
Secondly we present the change in stream observables $\vec{y}$ as a function of changes in model parameters $\vec{x}$, i.e., the derivative $d\vec{y}/d\vec{x}$ (\S\ref{sec:derivatives}).
Finally we present the adopted observational uncertainties, which set the covariance matrix $C_y$ (\S\ref{sec:datasets}).

\subsection{Model definition}
\label{sec:model}
The \CRLB\ formalism can only quantify information in the context of a model, which in our case is a model of a stellar stream in the gravitational potential of the Milky Way.
We consider cold stellar streams originating from disrupting globular clusters, which have been well-modeled with direct N-body simulations \citep[e.g.,][]{baumgardt2003, dehnen2004}.
Follow-up studies have shown that the phase-space distribution of the resulting debris is predominantly set by properties of the gravitational potential and the orbit of the progenitor, with a weaker dependence on the internal properties of the progenitor \citep{kupper2010, kupper2012}.
Hence, our model consists of parameters defining the gravitational potential and a 6-dimensional position of the progenitor.

To represent the global gravitational potential of the Milky Way, we use a combination of a Hernquist bulge \citep[parameterized with mass, $M_b$, and scale radius, $a_b$]{hernquist1990}, a Miyamoto-Nagai disk \citep[with parameters for disk mass, $M_d$, scale length, $a_d$, and scale height, $b_d$]{mn} and a spherical NFW halo \citep[parameterized with scale velocity, scale radius, and axis ratios $q_x=q_z=1$]{nfw}.
We provide the fiducial values for this model in Table~\ref{t:model}, along with measurement uncertainties where available.
This potential is similar to \texttt{MWPotential2014} \citep{galpy}, and fits a range of observed Milky Way properties. 

Next, we search for orbits of globular clusters in this fiducial gravitational potential that produce stellar streams similar to those observed in the Milky Way.
Our final sample contains analogs of ATLAS, GD-1, Hermus, Kwando, Orinoco, PS1A, PS1C, PS1D, PS1E, Sangarius and Triangulum streams, and in Appendix~\ref{sec:streams} we describe how we created streams in our fiducial gravitational potential that match properties of observed streams.
Briefly, we use the streakline method to forward-model stream observations in a modification of the \FF\ framework from \citet{bonaca2014}.
For each observed stream, we keep the potential fixed at fiducial values, and derive current progenitor positions, initial masses and stream ages that reproduce observations.
However, when calculating the \CRLB\ for streams, we fix the progenitor mass and age to their best-fitting values, as they are highly degenerate.
Instead, as model parameters we only use the present-day position of the progenitor, defined in the space of observables with the progenitor's two on-sky position angles ($RA_p$, $Dec_p$), distance $d_p$, radial velocity $V_{r,p}$ and two proper motion components ($\mu_{\alpha,p}$, $\mu_{\delta,p}$). 

The complete model for measuring the information content in stellar streams has 15 parameters: nine for the distribution of matter in the Galaxy and six for the position of the stream progenitor (as defined in Table~\ref{t:model}).
Some of these parameters have already been measured, and we include these prior constraints as appropriate. 
For example, progenitors of streams in our sample are unknown, so these are left completely unconstrained.
On the other hand, a substantial body of work has been dedicated to constraining the stellar components of the Milky Way.
As a result, properties of the bulge and the disk are known with a precision of $\lesssim10\%$ \citep{bobylev2017}.
For simplicity, we adopt $10\%$ priors, as indicated in Table~\ref{t:model}, and include them in covariance matrix $V$.
The halo mass, however, is still uncertain up to a factor of a few \citep[e.g.,][]{eadie2016, zaritsky2017}, and the reported constraints on the halo shape are conflicting \citep[e.g.,][]{loebman2014, bowden2016}, so we measure the information that streams provide on the dark matter halo independently of prior work.

\begin{table}
\begin{center}
\begin{tabular}{l c c c l}
\hline
\hline
Parameter & Symbol & Fiducial value & Uncertainty \\
\hline
\hline
Bulge mass & $M_b$ & $5\times10^{9}\rm M_\odot$ & $0.5\times10^{9}\rm M_\odot$ \\
Bulge scale radius & $a_b$ & 0.7\,kpc & 0.07\,kpc \\
Disk mass & $M_d$ & $6.8\times10^{10}\rm M_\odot$ & $0.68\times10^{10}\rm M_\odot$ \\
Disk scale radius & $a_d$ & 3\,kpc & 0.3\,kpc\\
Disk scale height & $b_d$ & 0.28\,kpc& 0.028\,kpc \\
\hline
Halo scale velocity & $V_h$ & 430\,km\,s$^{-1}$ & N/A \\
Halo scale radius & $R_h$ & 30\,kpc & --''-- \\
Halo $y$ axis ratio & $q_y$ & 1 & --''-- \\
Halo $z$ axis ratio & $q_z$ & 1 & --''-- \\
\hline
Progenitor RA & $RA_p$ & varied & N/A \\
Progenitor Dec & $Dec_p$ & --''-- & --''-- \\
Progenitor distance & $d_p$ & --''-- & --''-- \\
Progenitor radial velocity & $V_{r,p}$ & --''-- & --''-- \\
Progenitor RA proper motion & $\mu_{\alpha,p}$ & --''-- & --''-- \\
Progenitor Dec proper motion & $\mu_{\delta,p}$ & --''-- & --''-- \\
\hline
\hline
\end{tabular}
\caption{Model parameters}
\label{t:model}
\end{center}
\end{table}

\subsection{Calculating numerical derivatives for the \CRLB}
\label{sec:derivatives}
Intuitively, we can think of the Cram\'er--Rao bounds as quantifying how much we can change the parameters of a model $\vec{x}$, without violating the observational uncertainties of our data $\vec{y}$.
So, to calculate these bounds, we need to know how much the observed quantities $\vec{y}$ vary as a function of model parameters $\vec{x}$, or formally the derivative $d\vec{y}/d\vec{x}$.
Given that in our case the data are a collection of points in a 6-dimensional space (3D positions and 3D velocities of stream members), this derivative becomes a non-trivial calculation.
In what follows, we describe how we measure differences in stream models with different input parameters.

\begin{figure}
\begin{center}
\includegraphics[width=0.8\textwidth]{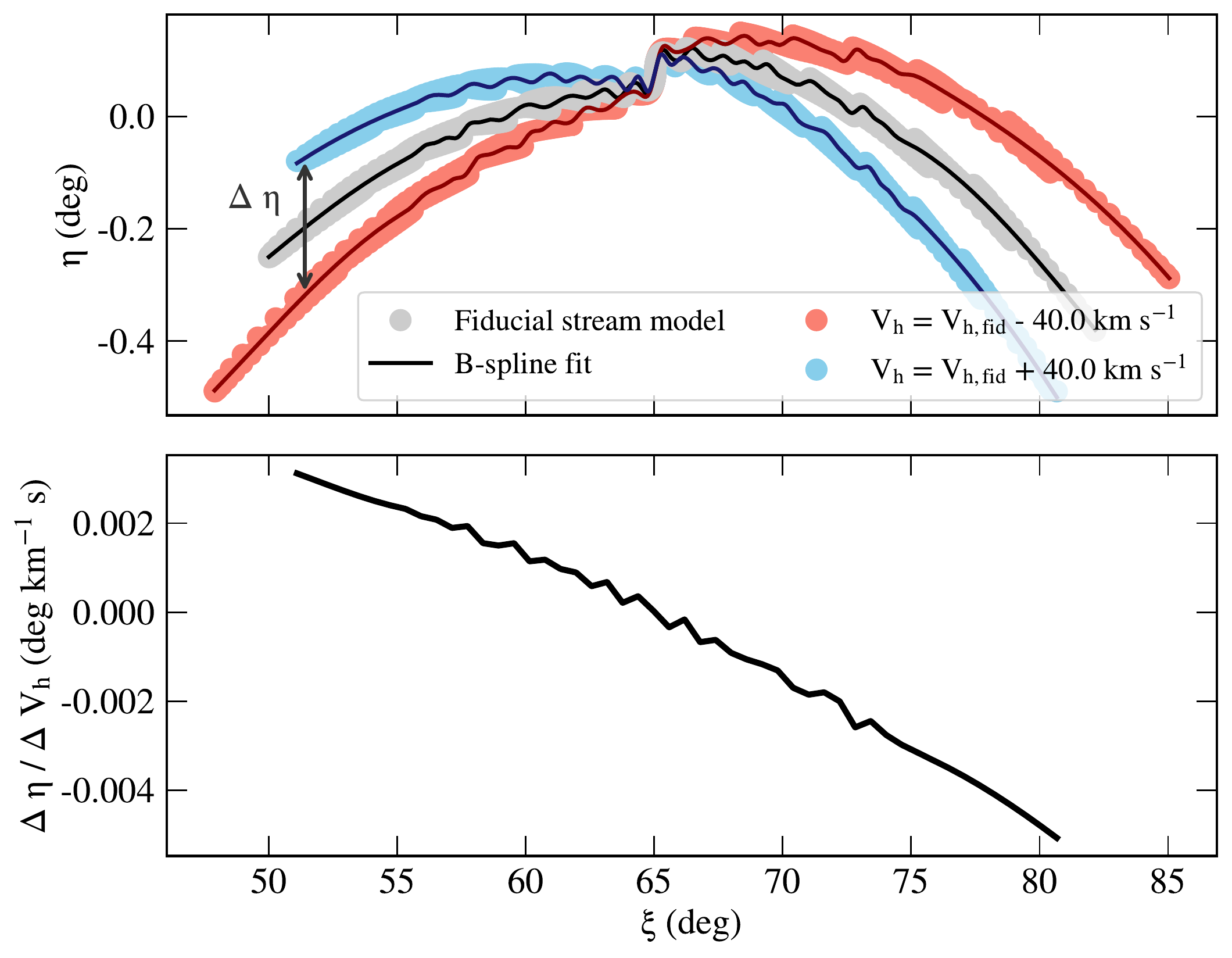}
\caption{We quantify the response of streams on changes in model parameters by calculating numerical derivatives, and in this figure we visualize the $d\eta / d V_h$ derivative (change in the sky position with the change in scale velocity of the underlying dark matter halo).
(\emph{Top}) The on-sky positions of the fiducial stream are shown in gray points, while the colored points show models in a less massive (red) and more massive halo (blue).
Solid lines show B-spline fits through these stream models.
The coordinate system is rotated such that the $\xi$ coordinate is approximately along the fiducial stream model, and $\eta$ is perpendicular to this stream track. 
(\emph{Bottom}) We define the derivative $d\eta / d V_h$ as the numerical derivative $\Delta\eta / \Delta V_h$.
At a fixed coordinate $\xi$ along the stream, the numerical derivative is the ratio of the difference between $\eta$ coordinates of models in the more massive and the less massive halo ($\Delta\eta$, marked with arrows in the top panel), and the total difference in halo scale velocity between these two models ($\Delta V_h$, 80\,km\,s$^{-1}$ in this example).
This numerical derivative is shown on the bottom panel as a function of the position along the stream.
}
\label{fig:derivative_steps}
\end{center}
\end{figure}

Cold streams analyzed in this work are very thin, most of them being at least two orders of magnitude longer than they are wide, and we treat them as one dimensional in the plane of the sky.
With each stream, we work in a spherical coordinate system ($\xi$, $\eta$) whose equator ($\eta=0$) is a great circle best-fitting the stream track, and the $\xi$ coordinate is our independent variable.
An example of this transformation is shown in Figure~\ref{fig:derivative_steps}, which shows positions of ATLAS-like stream members in a rotated coordinate system.

We measure deviations from the fiducial model at fixed positions along the stream ($\xi$), and use these numerical derivatives in our \CRLB\ calculation.
We illustrate the calculation of the $d\eta/d V_h$ derivative in Figure~\ref{fig:derivative_steps}, where in the top we show the on-sky position of a fiducial stream model in gray, and in red and blue models with a lower and higher halo scale velocity, respectively.
The numerical derivative at a position $\xi_k$ is then simply the difference between the $\eta(\xi_k)$ positions in the models of different scale velocity, divided by the difference in scale velocity between the models.
The bottom panel of Figure~\ref{fig:derivative_steps} shows how this derivative varies along the stream.
Derivatives in other data dimensions (distance, radial velocity, proper motions), and with respect to other model parameters, are calculated in the same fashion.
More formally, we calculate numerical derivatives using the following expression:
\begin{equation}
\left.\frac{d y_i}{d x_j}\right\rvert_{\xi_k} = \left.\left(\frac{y_i(x_{0,j} + \Delta x_j) - y_i(x_{0,j} - \Delta x_j)}{2\Delta x_j}\right)\right\rvert_{\xi_k}
\label{eq:derivative}
\end{equation}
where $y_i$ is the observable (either position $\eta$, distance, or one of the observed velocity components), $x_{0,j}$ is the fiducial value of parameter $x_j$, $\Delta x_j$ is a small value, and $\xi_k$ is the position along the stream where the derivative is evaluated.
A point to note is that we are not sensitive to changes along the stream within this formalism, which effectively reduces the dimensionality of our data, and might appear as not using the data to its full extent.
However, the coordinate along the stream (i.e., the stream length) is extremely correlated with the stream age, which is very poorly constrained with just the phase space distribution of the debris.
Consequently, not considering variations along the stream results in only a negligible loss of information.

\begin{figure*}
\begin{center}
\includegraphics[width=\textwidth]{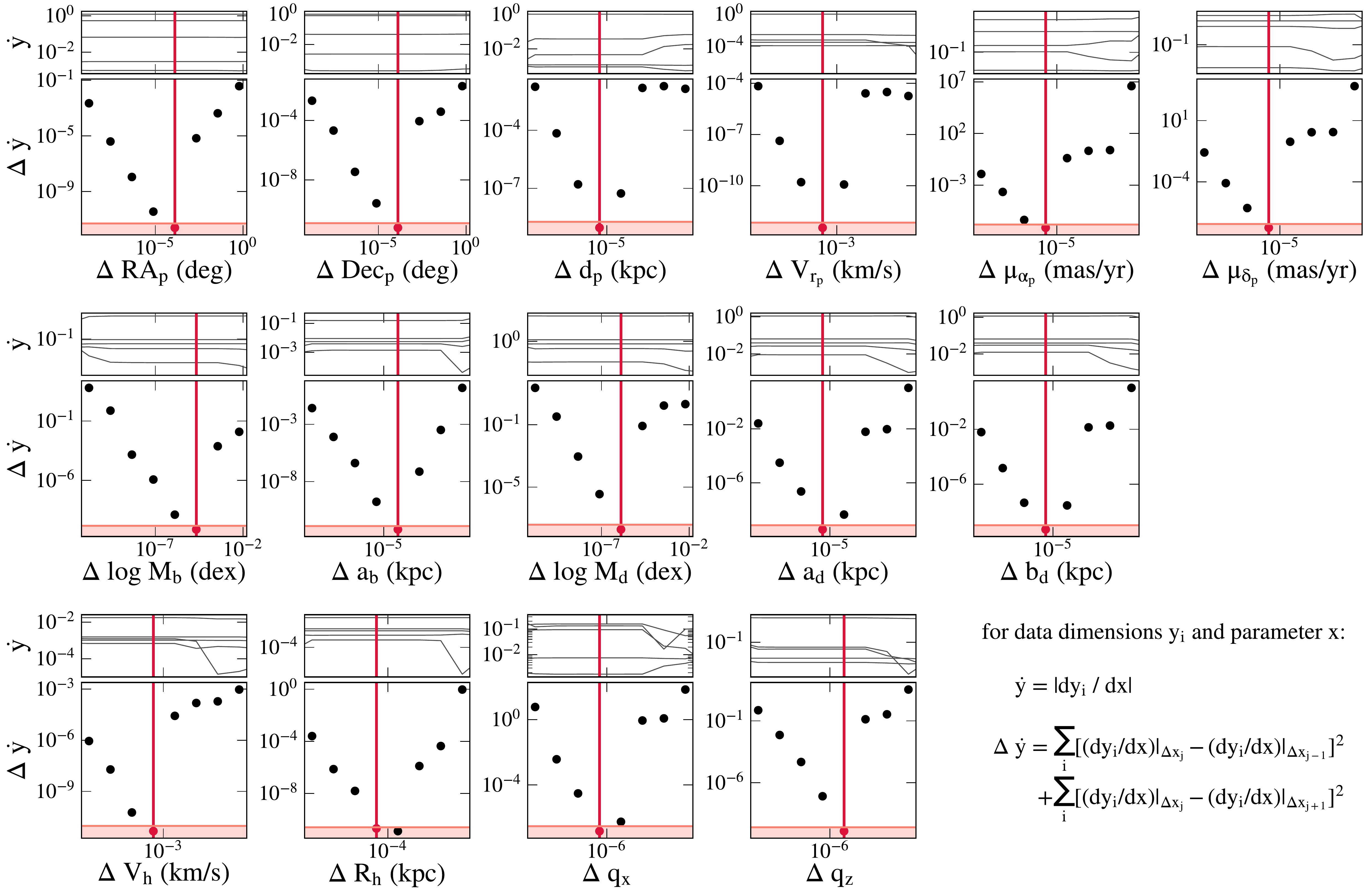}
\caption{Optimal steps for calculating numerical derivatives are marked with red vertical lines for all the parameters in our model of an ATLAS-like stream.
The top two rows present progenitor parameters, starting with progenitor's position, $RA_p$, on the left, through its proper motion $\mu_{\delta,p}$ on the right (for full description of model parameters, see Table~\ref{t:model}).
The first of these two rows shows the values of evaluated numerical derivative, $\Delta y/\Delta x$, as a function of step size $\Delta x$, and the five different lines are for each of the stream observables. 
The second row shows how the derivatives evaluated at a step size $\Delta x_j$ deviate from derivatives at adjacent step sizes $\Delta x_{j-1}$ and $\Delta x_{j+1}$.
Similarly, the middle two rows present derivatives for parameters defining the bulge and the disk, while the bottom two rows are dedicated to parameters of the dark matter halo.
For all of the model parameters, there is a local minimum in derivative deviation where the derivatives are the most stable.
We adopt the optimal step size (red points and vertical lines) as the smallest step with a derivative deviation within a factor of 2 from the local minimum.
}
\label{fig:derivative_conv}
\end{center}
\end{figure*}

The above prescription for calculating numerical derivatives relies (1) on the ability to evaluate any stream observable at a given position along the stream $\xi$, and (2) on the size of the parameter step $\Delta x$, and we now describe our strategies for addressing these points.
Both physical and model streams are a collection of stars, and thus all of the observables are discrete in nature (points in Figure~\ref{fig:derivative_steps}), which complicates comparison of different stream models.
However, to the first order, streams are one dimensional structures, and lines in the top panel of Figure~\ref{fig:derivative_steps} show that a B-spline fits well the distribution of stream positions.
We proceed by representing a stream model with a B-spline, which makes for a trivial evaluation of a stream observable at any position $\xi_k$.
This approach only takes into account information provided by the mean stream track, but not density variations along the track, and as such does not estimate the full information content in streams.
Nevertheless, the observed density variations along the streams are a convolution of true variations and observational incompleteness, which typically remains unconstrained.
Since the measurements of stream densities are uncertain, and we expect most of the information on the gravitational potential to be encoded in the stream track, in this work we focus on analyzing only the information content in the average positions of stream members, so approximating a stream model with a line is appropriate.

Finally, to robustly evaluate numerical derivatives entering the calculation of Cram\' er--Rao bounds, we explore how the derivatives depend on $\Delta x$, the parameter step size from Equation~\ref{eq:derivative}.
To find the optimal $\Delta x$, we vary the step size by 10 orders of magnitude for every model parameter, store values of derivatives for different observables at several positions along the stream, and search for the step size where the derivatives are the most stable.
Our metric for derivative stability at a step size $\Delta x_j$ is the overall deviation of derivatives evaluated at the adjacent step sizes $\Delta x_{j-1}$ and $\Delta x_{j+1}$, denoted $\Delta \dot{y}$ and defined as:
\begin{equation}
\Delta \dot{y} = \sum_i \left(\left.\frac{dy_i}{dx}\right\vert_{\Delta x_j} - \left.\frac{dy_i}{dx}\right\vert_{\Delta x_{j-1}}\right)^2 + \sum_i \left(\left.\frac{dy_i}{dx}\right\vert_{\Delta x_j} - \left.\frac{dy_i}{dx}\right\vert_{\Delta x_{j+1}}\right)^2
\label{eq:stability}
\end{equation}
where $y_i$ are observables (positions, distances and kinematics along the stream).
In Figure~\ref{fig:derivative_conv} we show numerical derivatives (short panels) and derivative deviations (square panels) as a function of parameter step size for the ATLAS-like stream in our sample.
Each row is dedicated to a group of parameters, with progenitor position on the top, parameters of the baryonic components in the middle, and parameters describing the dark matter halo in the bottom row.
At extremely large or small steps, the derivatives are unstable, which is evident as the derivatives themselves change (short panels), and also from their large deviations (square panels).
For all parameters, however, at intermediate step sizes the derivatives are constant and their deviations are small.
We adopt the smallest step size with a deviation within a factor of 2 from the minimum value (red point and vertical line) for calculating numerical derivatives.
Large deviations for large step sizes are due to derivatives being in the non-linear regime, while large deviations at small steps are reflecting the numerical noise.
The adopted step size is in between these two regimes, and is calculated for each stream individually.
For most parameters, the adopted size is rather small (e.g., on the order of $\rm m\,s^{-1}$ for velocities), but it still produces physically different stream models, underlying the importance of quantitative selection of step size in numerical differentiation.

While we presented a robust method to calculate numerical derivatives, a fully self-consistent solution would be to obtain derivatives directly when creating a stream model.
This is usually done using auto-differentiation, which has not been implemented in our legacy orbit integrator.
Evaluating exact derivatives of stream models with respect to input model parameters remains the most important technical improvement left for future work, and is discussed in \S\,\ref{sec:dis_extensions}.

\subsection{Sets of observational data}
\label{sec:datasets}

The amount of information we can learn about the gravitational potential from a stream data set depends both on the intrinsic sensitivity of a stream on different aspects of the gravitational potential, and also on the properties of the data set itself.
The intrinsic sensitivity, encoded in the derivatives of stream observables with respect to potential parameters, has been discussed above.
In this section, we set up different stream data sets, defined by the type, amount and precision of observations at hand.

For simplicity, we assume that observations across the different data sets are uniformly distributed every 0.5\,deg along each stream, with a minimum of 15 measurement points per stream.
We also assume that all the different types of data (e.g., radial velocity, proper motions) are measured at these same positions along the stream and that the uncertainties are the same for a given type of observation.
All of these assumptions make for a highly idealized scenario: there are density inhomogeneities along the streams, so the likely members are hardly equidistant; different types of data are typically obtained for different stars; and observational uncertainties are a function of stellar brightness, color, and often distance.
However, in this work we merely aim to demonstrate a framework for calculating the information content of a stream data set.
More realistic data can be analyzed in the same way to produce tailored predictions.

\begin{table}
\begin{center}
\begin{tabular}{l c c c c c c}
\hline
\hline
Data set & $\sigma_\eta$ (deg) & $\sigma_d$ (kpc) & $\sigma_{V_r}$ (km\,s$^{-1}$) & $\sigma_{\mu_\alpha\star}$ (mas\,yr$^{-1}$) & $\sigma_{\mu_\delta}$ (mas\,yr$^{-1}$) \\
\hline
Fiducial & 0.1 & 2.0 & 5 & 0.1 & 0.1 \\
DESI & 0.1 & 2.0 & 10 & N/A & N/A \\
Gaia & 0.1 & 0.2 & 10 & 0.2 & 0.2 \\

\hline
\hline
\end{tabular}
\caption{Observational data sets}
\label{t:datasets}
\end{center}
\end{table}

In measuring the information content in the Milky Way streams, we consider the following scenarios of data availability:
\begin{enumerate}
\item fiducial data set obtainable from wide-field photometric surveys with targeted kinematic follow-up,
\item present data on positions of stream stars with radial velocities from DESI and DESI-like surveys, and
\item data on stream members from the final data release of the Gaia mission.
\end{enumerate}
Adopted observational uncertainties for these scenarios are summarized in Table~\ref{t:datasets}, and justified in more detail below.

For the fiducial data set, we assume that the positions of stream members come from photometric surveys, radial velocities from targeted spectroscopic follow-up, and proper motions from long-baseline, spacecraft observations.
The on-sky positions of stars are measured very precisely \citep[for example, better than 100\,mas in SDSS,][]{pier2003}, so the uncertainty in stream position equals the stream width, rather than the uncertainty in positions of stars themselves.
Globular clusters produce thin streams, with widths between $\sim$0.1\,deg \citep[e.g., PS1A,][]{bernard2016} and $\sim$0.4\,deg \citep[e.g., Sangarius,][]{grillmair2017a}.
We adopt 0.1\,deg for our fiducial positional uncertainty.
Unlike the positions projected on the sky, the distances to most of the streams are highly uncertain, as there are typically no standard candles identified along the stream.
The main distance indicator is merely the position of the main sequence turn-off \citep[e.g.,][]{bonaca2012}, which provides a distance uncertainty of $\sim20\,\%$ under the assumption that the age and metallicity of the stream stars are known.
For simplicity, we adopt a single value of $2\,\rm kpc$ as our fiducial uncertainty in distance.
This value is conservative for nearby streams, but likely overly optimistic for the more distant ones.
On the kinematics side, our fiducial uncertainty for radial velocities is conservatively 5\,km\,s$^{-1}$, as a number of medium-resolution spectrographs have demonstrated performance on the level of a few km\,s$^{-1}$ for individual stars \citep[e.g.,][]{sg}.
And finally, space-based astrometry with a long baseline allows proper motions to be measured with an uncertainty of only 0.1\,mas\,yr$^{-1}$ \citep{sohn2015}, which we also use as our fiducial uncertainty.
These uncertainties are among the best attainable at the present.
Most of the streams, however, only have positional data, and only a few have kinematic information from radial velocities.
So within the fiducial case, we also consider separate situations of having access to 3D (positions only), 4D (positions and radial velocities) and 6D data (full phase space).

In the second case, we emulate a data set whose positions originate from photometric surveys, same as in the fiducial case, but which includes radial velocities from lower-resolution spectroscopic surveys.
We study this scenario in anticipation of a number of highly multiplexed, $R\approx5,000$ surveys being launched in the next few years \citep[e.g., DESI,][]{desi}.
These surveys were designed to target objects as faint as $g=20$, and are expected to deliver radial velocities for the Milky Way sources with an uncertainty of 10\,km\,s$^{-1}$, which we adopt for this case.
Albeit this value is a factor of two worse than the radial velocity precision in our fiducial case, these surveys will operate very quickly, and have the potential to provide kinematics for all of the known streams -- a feat hard to accomplish with targeted follow-up.
With this scenario, we will specifically explore the possibility of measuring the gravitational potential with a large number of streams with medium velocity precision.

Our final case studies data in the post-Gaia era, where distances come from Gaia parallaxes, radial velocities from Gaia's Radial Velocity Spectrometer (RVS), and proper motions from the 5-year Gaia astrometry.
The uncertainty in stream position will still be limited by its intrinsic width, but the distance estimates should be improved by an order of magnitude with respect to the fiducial case.
Typical Gaia parallax uncertainty for bright red giants will be 0.01\,mas\,yr$^{-1}$ at the end of the nominal mission \citep{perryman2001, prusti2016}, which translates to a distance uncertainty of 1\,kpc at 10\,kpc.
Distance precision is expected to increase by a factor of 5 after matching the astrophysical parameters of these stars \citep{mcmillan2017, ting2018}, so we use 0.2\,kpc as our distance uncertainty in this case.
For a similar population of stars, RVS will deliver radial velocities precise to 10\,km\,s$^{-1}$.
Lastly, proper motion precision is $\approx0.2$\,mas\,yr$^{-1}$ for faint, blue stars, such as those at the main sequence turn-off of stellar streams.
We adopt this value for our proper motion uncertainty, as we expect faint blue stars to make up the majority of identified stream members.
Data in this scenario are assumed to span the full phase space, and are distinguished from the fiducial 6D case only by the expected measurement uncertainties.
Distances are slightly more precise, while the kinematics are a factor of 2 worse than in the fiducial case, so when comparing the two, we will be able to gauge relative importance of different types of data for dynamical modeling of streams.

\section{Results}
\label{sec:results}
Following the steps outlined in section~\S\,\ref{sec:method}, we inferred the information content in 11 Milky Way-like streams.
In this section, we first present constraints on the underlying gravitational potential from a single stream (\S\,\ref{sec:res_ind}), then compare constraints from different streams (\S\,\ref{sec:res_comp}), and finally explore joint constraints from multiple streams (\S\,\ref{sec:res_joint}).

\subsection{Constraints from an individual stream}
\label{sec:res_ind}

\begin{figure*}
\begin{center}
\includegraphics[width=\textwidth]{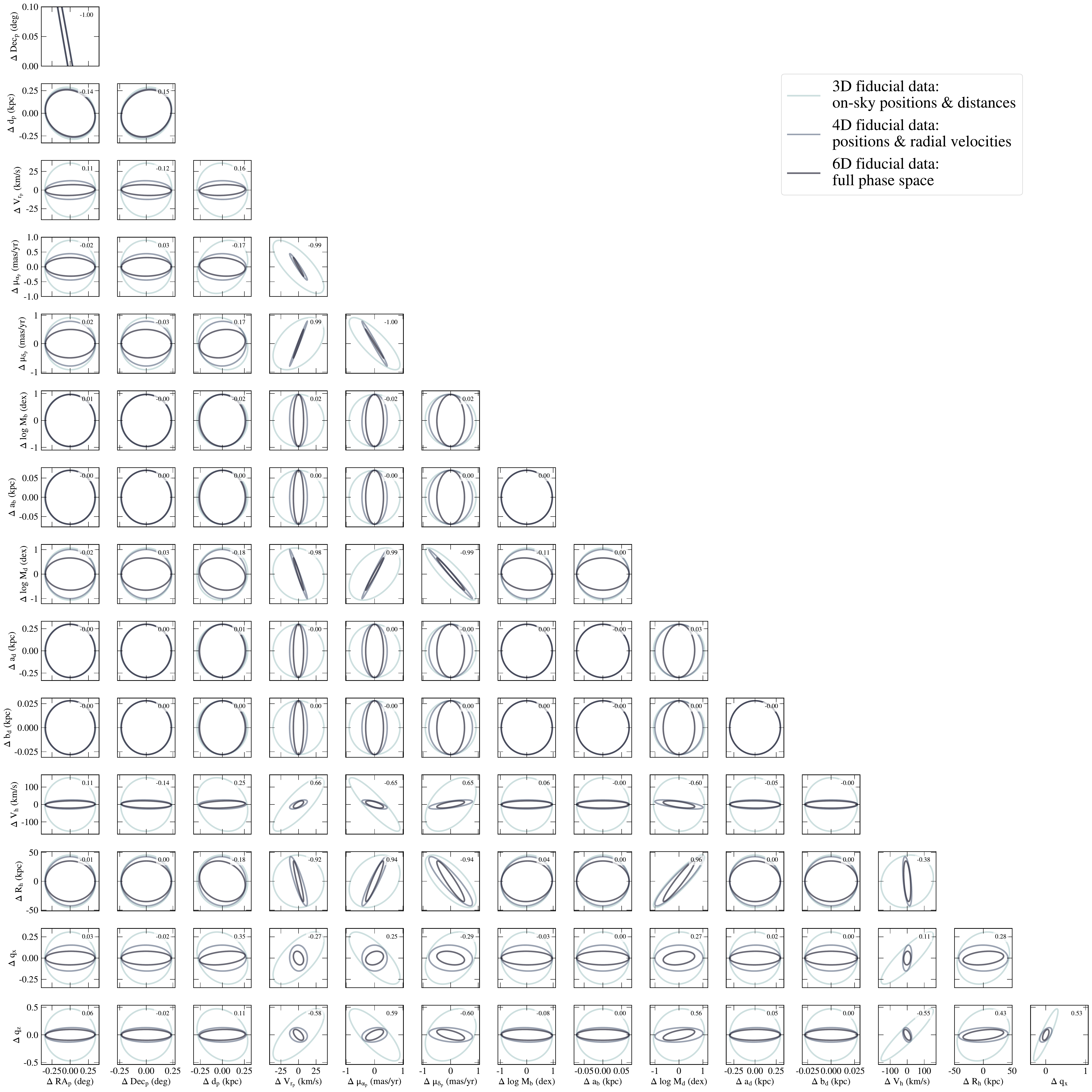}
\caption{A visualization of correlations between Cram\'er--Rao bounds (ellipses) on all 15 of the model parameters for an ATLAS-like stream.
The lightest colored ellipses show constraints achievable when only on-sky positions and distances are known, medium shaded ellipses are constraints produced with an addition of radial velocities, while the darkest ellipses come from the full 6D phase space information on stream members.
The addition of kinematic data drastically improves constraints on most of the parameters, indicating that kinematics are essential for meaningful recovery of the Galactic potential.
However, streams seem to contain little information on the baryonic matter components of the Galaxy -- these constraints do not improve with more data and remain prior-dominated.
}
\label{fig:crb_correlations}
\end{center}
\end{figure*}

We estimate how precisely every stream data set constrains the current position of the stream's progenitor, as well as the underlying gravitational potential.
For a given combination of a stream and an observational setup, all of this information is contained in a matrix of Cram\'er--Rao lower bounds (\CRLB, $C_x$).
In Section~\ref{sec:method}, we showed how to calculate its inverse -- the Fisher matrix $C_x^{-1}$ (Equation~\ref{eq:crlb}).
If the condition number of $C_x^{-1}$ is small, it can be easily inverted to obtain the Cram\'er--Rao lower bounds.
However, when the model parameters are poorly constrained, $C_x^{-1}$ has a large condition number, and it is then non-invertible by standard libraries.
We find that condition numbers of calculated $C_x^{-1}$ tend to be large, so to get the inverse $C_x$, we use a robust, iterative method of matrix inversion, described in Appendix~\ref{sec:inversion}.

Once we have the \CRLB, the square root of its diagonal are the bounds on individual parameters of our model.
If all of the model parameters were uncorrelated, the \CRLB\ matrix would be diagonal.
However, we find that \CRLB\ have off-diagonal elements for all of the streams, and in the remainder of this section, we discuss covariances between different model parameters inferred from observations of an ATLAS-like stream.

The corner plot in Figure~\ref{fig:crb_correlations} visualizes the structure of the \CRLB\ matrix for an ATLAS-like stream, under the assumption of fiducial observational uncertainties.
Ellipses in each panel show covariant CRLB for a single pair of model parameters, with different colors indicating the dimension of the assumed fiducial data set: the lightest ellipses are constraints from 3D positions only, the medium ellipses include 3D positions and radial velocities, and the darkest ellipses are for the full 6D fiducial data set.
Properties of an ellipse for a pair of parameters $i,j$ are determined from a slice of the Fisher matrix $C_x$ such that its rotation angle $\theta$, width $w$ and height $h$ are given by:
\begin{align*}
\theta &= {\rm arc tan}\left(\frac{V_{1,0}}{V_{0,0}}\right) \\
w &= 2\,\sqrt{v_0} \\
h &= 2\,\sqrt{v_1}
\end{align*}
where $V_k$ are the eigenvectors ($V_{k,l}$ is the $l$ component of the eigenvector $k$) and $v_k$ the corresponding eigenvalues of the Fisher-matrix slice $M_{ij} \equiv \left[\left[C_{x,ii}, C_{x,ij}\right], \left[C_{x,ji}, C_{x,jj}\right]\right]$.

In general, additional stream data leads to tighter constraints on model parameters, and consequently darker ellipses in Figure~\ref{fig:crb_correlations} are smaller than lighter ellipses, although the relative sizes vary between different pairs of parameters.
In some cases, the impact of additional data is drastic.
For example, the presence of radial velocities in a data set collapses the uncertainty in the progenitor radial velocity, $V_{r_p}$, by an order of magnitude (compare the lightest and medium-colored ellipses in the fourth column from the left in Figure~\ref{fig:crb_correlations}).
For other parameters, additional data can break degeneracies existing in bounds from less extensive data sets, such as that between the halo scale velocity, $V_h$, and halo $x$-axis ratio, $q_x$, which are tightly correlated when constrained by a 3D data set, but show hardly any correlation when analyzed with 6D data (see panel in the second row from the bottom, and the third column from the right).
Typically however, additional data only tightens parameter constraints, and keeps the correlations between parameters in place.
Finally, for some parameters, additional data provide almost no additional constraints.
For the ATLAS-like stream presented in Figure~\ref{fig:crb_correlations}, additional data have no impact on the on-sky position of the progenitor, which is mainly inferred from the on-sky positions of stream stars, as well as on some bulge and disk properties, which are dominated by the prior.
Given that the pericenter of this stream is 23\,kpc, which is approximately 30 and 7 times beyond the bulge and disk scale lengths, respectively, it is unsurprising that its debris provides little information on these components.
This example serves a cautionary tale -- obtaining additional data along a stream can be expensive, so it is important to quantify the expected gains beforehand, and we have provided a framework to do so.

Data sets featuring only positions of stream stars, provide weak, but not vanishing, constraints on model parameters.
This result is somewhat surprising, because in the absence of kinematics, the masses and timescales in the system can be arbitrarily rescaled to reproduce the same positions, so we expect that kinematic data is a prerequisite for any dynamical inferences.
However, we introduced a mass scale in the problem by fixing the mass of the stream progenitor, rather than using it as a model parameter.
This effectively assumes infinitely precise knowledge of the progenitor mass, which likely propagates into weak constraints on model parameters, as obtained in the absence of kinematic data.
Since these constraints from 3D data are much weaker than those from 4D and 6D data sets, we decided not to complicate our analysis by an introduction of another parameter, and simply note that they are driven by the prior knowledge of the progenitor properties.

As the final element of our exploration of a Fisher information matrix for a single stream, we discuss correlations between constraints on different parameters.
The CRLB ellipses in different panels of Figure~\ref{fig:crb_correlations} exhibit a whole range of correlations: from strongly and mildly (anti-)correlated, to those completely uncorrelated.
We quantify these covariances with the Pearson correlation coefficient, which for a pair of parameters $(i,j)$ is simply $p = C_{x,ij} / \sqrt{C_{x,ii}\,C_{x,jj}}$, where $C_{x,kl}$ is the $k,l$ element of the Fisher matrix $C_x$.
We report correlation coefficients for parameters constrained with 6D fiducial data set in the upper right corner of each panel in Figure~\ref{fig:crb_correlations}, and discuss below the most extreme values.

On-sky positions of an ATLAS-like progenitor are perfectly anti-correlated, which stems from the requirement that the position of the progenitor is on top of the one-dimensional stream track, and the stream's projected span from southeast to northwest.
Similarly, the progenitor's proper motions are anti-correlated, indicating that at least one component of the velocity vector is aligned with the stream.
This is further supported by the strong correlations between the progenitor's proper motions and its radial velocity -- a mark that there is a more physical decomposition of the progenitor's velocity vector.
One would expect similar behavior for the 3D spatial position of the progenitor, however, distances in our fiducial case are too uncertain to exhibit correlations with on-sky positions.

In addition to correlations arising from the geometry of stream observations, there are also covariances originating from the physical aspects of stream constraints on the gravitational potential.
For example, scale velocity and scale radius of the dark matter halo are mildly anti-correlated, which is expected if the stream is most sensitive to the total amount of matter within some volume of the Galaxy, but is somewhat agnostic to the way the matter is distributed.
In order to conserve the total mass in the halo, $M_h \propto V_h^2\,R_h$, the halo scale velocity needs to decrease as its scale radius increases, so the correlation between these two parameters is negative.
Similarly, the disk mass is mildly anti-correlated with the halo scale velocity, and strongly correlated with the halo scale radius.
This can again be understood in the context of streams only constraining the total mass in the inner galaxy, with contributions from both the disk and the halo.
The central halo density decreases as the halo scale radius increases (or the halo scale velocity decreases), so to conserve the total mass, the disk mass must also increase, which then introduces the observed correlations between the halo and disk parameters. 

There are also strong correlations between the halo scale radius and the progenitor's kinematics.
These underscore the importance of including the progenitor parameters in measuring the information content in stellar streams, as neglecting to do so would lead to overly optimistic estimates.
Furthermore, this mixing between parameters describing the progenitor and those describing the distribution of matter indicates that some streams might be at a more informative orbital phase or position relative to the Sun, a prospect we investigate in the next section where we compare stream performance in constraining parameters of the gravitational field.

\subsection{Comparison between streams}
\label{sec:res_comp}
In the context of our model, each stream provides information on the position of its individual progenitors, as well as on the distribution of matter in the Galaxy that they all have in common.
In this section, we discuss how constraints on various aspects of the Galactic gravitational field compare between 11 streams in our sample.

\begin{figure*}
\begin{center}
\includegraphics[width=\textwidth]{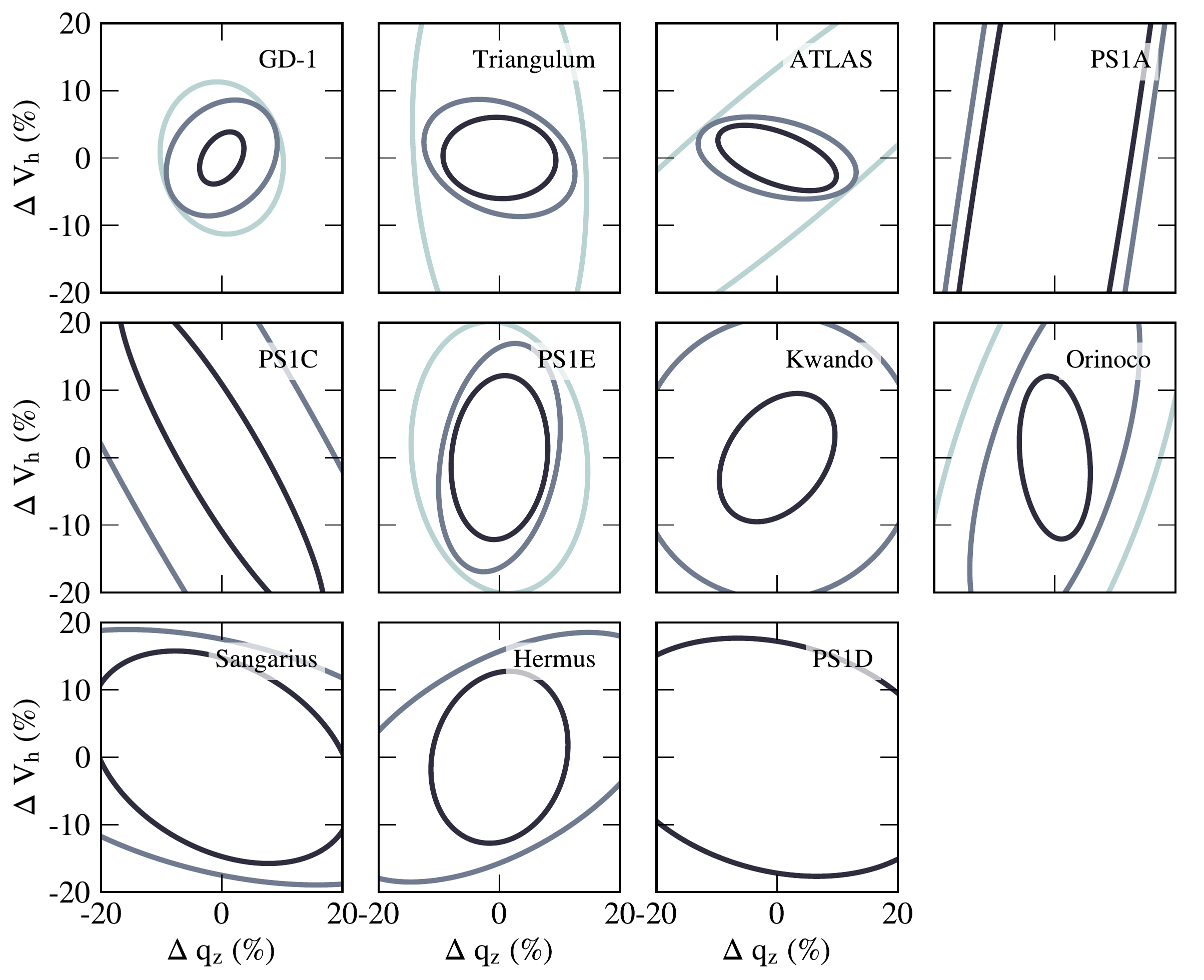}
\caption{Fractional Cram\'er--Rao bounds on the scale velocity and $z$-to-$y$ axis ratio of the dark matter halo, based on fiducial observations of 11 streams in our sample.
The lightest ellipses represent constraints from 3D spatial data, medium ellipses include information from radial velocities in addition to positions, and the darkest ellipses feature data sets with full 6D phase-space.
The quality of constraints varies among the streams from GD-1, where the constraints are within 20\% in every case considered, to PS1-D, for which only the 6D data set provides constraints better than 20\%.
The two halo parameters are differently correlated, so combining multiple streams will break some of these individual degeneracies.
}
\label{fig:crb2d_comparison}
\end{center}
\end{figure*}

In previous section, we explored correlations in Cram\' er--Rao bounds between all parameter pairs (Figure~\ref{fig:crb_correlations}), so we start this section by comparing different streams' constraints on a single pair of parameters.
Figure~\ref{fig:crb2d_comparison} features 2D Cram\' er--Rao bounds for the halo scale velocity and its $z$-axis flattening, with each panel in a grid dedicated to an individual stream.
Analogously to Figure~\ref{fig:crb_correlations}, light, medium through dark ellipses are constraints based on 3D, 4D and 6D fiducial data sets, respectively.
Unlike the previous figure, here the parameter constraints are relative, and capped at 20\,\% to highlight the most informative streams.
Performance of different streams in constraining halo scale velocity and halo shape varies from better than 20\,\% with only 3D data for streams such as GD-1 and PS1E, to worse than 20\,\% for PS1D even with 6D data, and everything in between.
Relative importance of the input data also varies among the streams; for example, additional radial velocities improve constraints from GD-1 only marginally, but they provide a major improvement in constraints from Triangulum and ATLAS.
An in-depth analysis of these bounds can inform an optimal observing strategy, and we explore implications that currently ongoing spectroscopic and astrometric surveys will have for tidal streams in Section~\S\,\ref{sec:forecast}.
Constraints on halo scale velocity and shape are correlated for all of the streams, but these correlations differ in both degree and direction.
In the following section, we investigate how combining different streams can break these degeneracies.
In the remainder of this section, we examine why different streams are differently sensitive to various model parameters, and study how \CRLB\ for individual parameters depend on properties of the stream and the orbit of its progenitor.

\begin{figure*}
\begin{center}
\includegraphics[width=\textwidth]{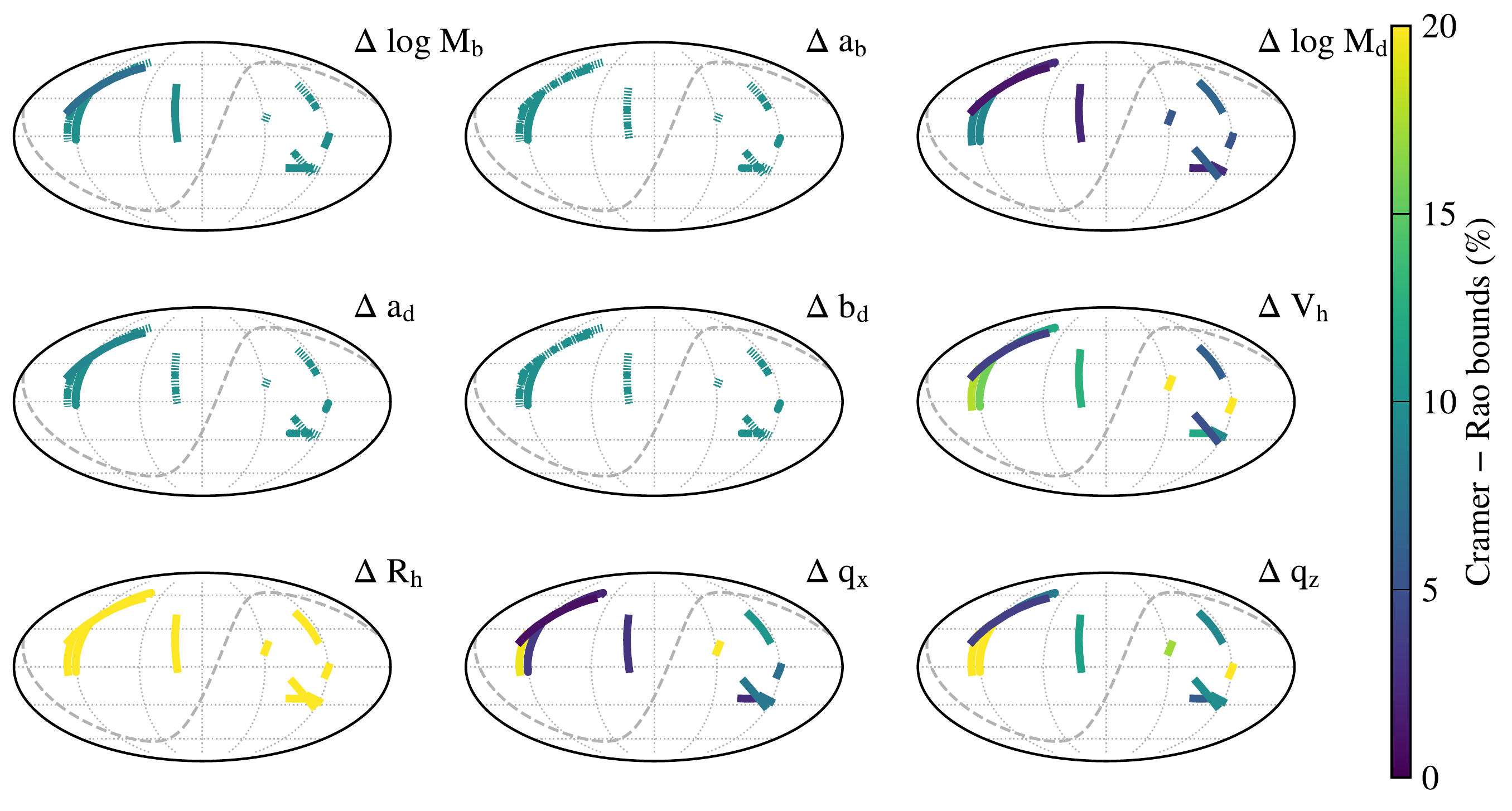}
\includegraphics[width=0.7\textwidth]{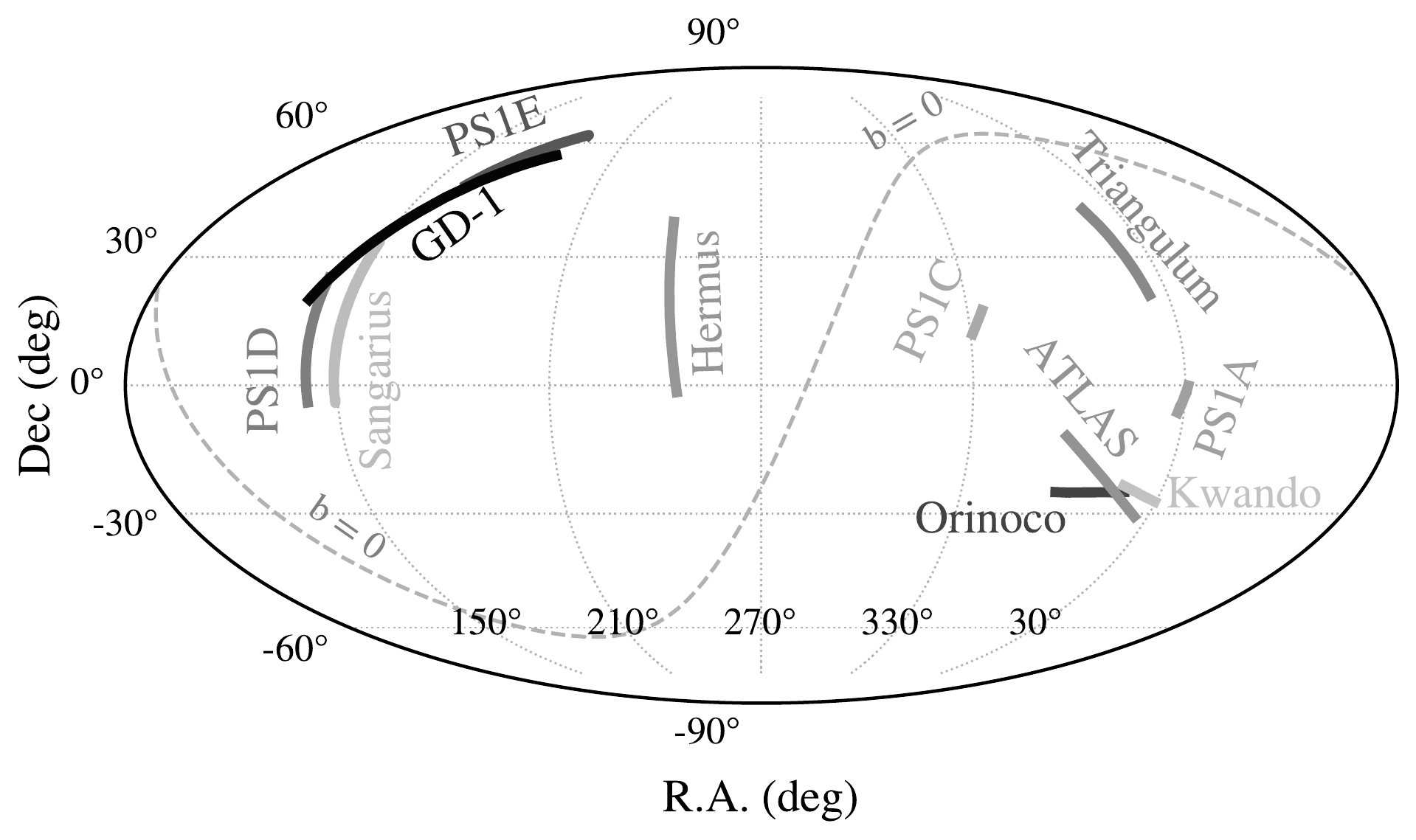}
\caption{Different streams are sensitive to different parameters of the Galactic potential.
On-sky positions of analyzed streams are color-coded by the relative precision expected from fiducial 6D observations, with each panel dedicated to a single parameter (labeled in the upper right corner).
Most of the baryonic parameters are set by the prior (streams shown as dotted lines), which is only improved upon for the disk mass (top right panel).
Constraints on the halo parameters, obtained with no prior information, are more diverse, so that even streams which appear close in the sky have very different constraining power.
This indicates that the origin of the information in streams goes beyond their current on-sky position.
}
\label{fig:sky_precision}
\end{center}
\end{figure*}

Figure~\ref{fig:sky_precision} summarizes \CRLB\ from fiducial 6D observations on all model parameters that streams in our sample have in common.
Each panel in a grid shows positions of streams on the sky (in equatorial coordinates), colored by the fractional CRLB on the parameter indicated in the upper right corner of the panel.
Similar to Figure~\ref{fig:crb2d_comparison}, the color-bar saturates at 20\,\% to highlight parameters for which streams provide interesting constraints.
The larger panel in the bottom has the scale and labels for each stream, and the dashed gray line marks the Galactic plane ($b=0$) on all panels.

In addition to stream observations, the bounds on baryonic parameters were informed by a 10\,\% prior, so the \CRLB\ on these parameters are better than 10\,\% by construction.
To identify parameters whose bounds are determined predominantly by prior information, we calculated Kullback--Leibler divergence (KLD) between the prior and the bound:
\begin{equation*}
D_{KL} = \int_{-\infty}^{\infty} p(x) \ln\frac{p(x)}{q(x)} dx
\end{equation*}
where $p(x) = \mathcal{N}(0,\sigma_{bound})$ is the normal distribution centered on zero, with the parameter's \CRLB\ as a dispersion, and similarly $q(x) = \mathcal{N}(0,\sigma_{prior})$ is the normal distribution used as a prior on that parameter in the calculation of the \CRLB.
In general, KLD compares how similar two distributions are, so in our case, a small divergence means that the bound contains little information from streams.
Quantitatively, we call a parameter constraint prior-driven if its KLD with respect to the prior is smaller than $D_{KL}<0.01$, and plot these bounds with dotted lines in Figure~\ref{fig:sky_precision}.
None of the streams improved upon the prior information for the length scale of the bulge and disk, only several have improved on the bulge mass measurement (the inner halo streams GD-1, Hermus, Orinoco, and PS1A), while all the streams have constrained the disk mass further.
Mock streams in our sample are sensitive to the overall mass in the baryonic components, but not to its spatial distribution, which is likely a result of them never reaching closer than 4\,kpc from the Galactic center.
In the rest of this section, we focus on parameters for which streams provide competitive constraints: the mass of the disk, and the halo parameters.

Across the different parameters, streams provide diverse constraints: some parameters are constrained to the same degree by all of the streams, while there is variance from a few percent to more than 20\,\% for others.
For example, none of the streams constrain halo scale radius better than 20\,\%, which is not entirely unexpected for this sample of streams where only four venture past the break in the NFW profile at 30\,kpc.
The constraints for other parameters are more varied across the sample; some streams like GD-1 are good at constraining all parameters, while some have a mediocre overall performance, but excel in one parameter, such as Sangarius for the halo $x$-axis flattening, $q_x$.
There are no obvious trends between the quality of constraints and the stream position on the sky -- rather, streams observed in the similar region of the sky can have very different sensitivity to halo parameters (e.g., PS1D is nearly parallel and close to Sangarius, but its constraints on $q_x$ are worse than 20\,\%).

\begin{figure*}
\begin{center}
\includegraphics[width=\textwidth]{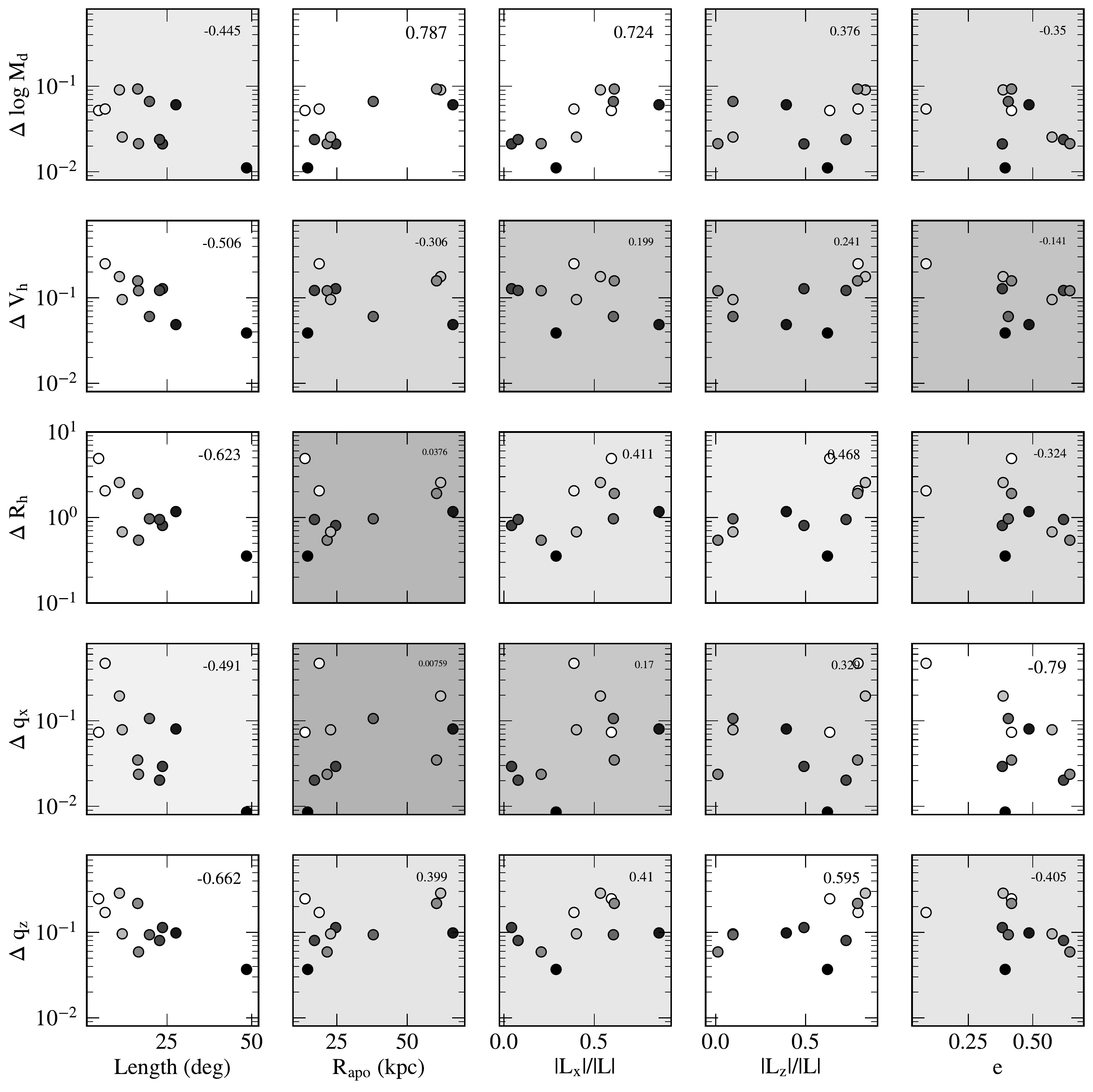}
\caption{Dependence of the precision in recovery of different model parameters (rows, from top: disk mass, halo scale velocity, halo scale radius, halo $x$ and $z$-axis flattening) as a function of different stream properties (columns, from left: stream length, apocentric radius, median $x$ and $z$-components of the angular momentum).
The correlation in every panel is indicated in the top right corner with a $p$~value and with the panel brightness.
The most robust trend is with stream length, as long streams in general produce more precise constraints.
The disk mass is better constrained by streams whose angular momentum vector is in the disk plane, while streams on eccentric orbits are more sensitive to the halo shape.}
\label{fig:orbit_correlations}
\end{center}
\end{figure*}

We next test the dependence of stream sensitivity to parameters of the gravitational potential on the intrinsic properties of the system, i.e. the orbital properties of the progenitor.
Specifically, Figure~\ref{fig:orbit_correlations} shows how the \CRLB\ on the disk mass and halo parameters (y axis) depend on stream length, apocentric radius, orbital orientation and eccentricity (x axis).
The grid is organized such that rows share the same model parameters, while columns share the same stream property.
Each point in a panel represents a mock stream, color-coded by its length.
The Pearson correlation coefficient ($p$~value) between the model parameter and the stream property is given in the top right corner of every panel.
To emphasize stronger correlations, the text size of the quoted $p$~value and the brightness of the panel's background increase with the absolute size of the $p$~value.

The Cram\' er--Rao bounds on all model parameters are better for longer streams.
For most parameters, this correlation with stream length is stronger than with any other stream property.
Differences in stream models in halos of different halo scale velocity, as illustrated in Figure~\ref{fig:derivative_steps}, are easier to discern at the ends of a stream.
Given a fixed observational uncertainty, this means that longer streams should in general be more sensitive to changes in the gravitational potential -- a hypothesis quantitatively confirmed with \CRLB\ of all model parameters improving with stream length.

The length of a stream is its only intrinsic property which universally correlates with the \CRLB\ of model parameters, but other features of streams can be linked to a specific parameter of our galaxy model.
For example, streams with smaller apocenters constrain the disk mass better (as visible in the top row, and second panel from the left of Figure~\ref{fig:orbit_correlations}).
These streams never venture out in the halo, so their gravitational potential is dominated by the disk, and it is consequently the mass of the disk they pinpoint.
Also intuitively, the flattening of the halo along the $z$-axis is better constrained by streams whose orbital planes are perpendicular to the Galactic disk.
The $z$ component of the angular momentum is small on such orbits, and in the fourth panel of the bottom row of Figure~\ref{fig:orbit_correlations} we indeed see that the best \CRLB\ are obtained for streams with low value of $|L_z|/|L|$, where $L$ is the angular momentum averaged along the orbit.
And finally, \CRLB\ on several model parameters anti-correlate at varying degrees with the eccentricity of the progenitor's orbit.
Larger eccentricity means that the stream has orbited in a larger fraction of the Galaxy's volume, which seems to particularly inform parameters on the halo shape, such as the axis ratios and the scale radius.
At a fixed eccentricity, longer streams provide better constraints, which is especially evident for a number of streams with an eccentricity of $e\approx0.4$.
Given the complex dependence of \CRLB\ on different stream properties, we can expect an interesting interplay between these drivers when combining information from multiple streams.

\subsection{Joint constraints}
\label{sec:res_joint}
So far, we have only analyzed the information content in individual streams.
However, streams in our sample are independent experiments, so information they provide on the gravitational potential can be trivially combined.
In this section we first show how to calculate joint constraints from multiple streams, and then discuss resulting improvements in inferred properties of the Galactic dark matter halo that all streams have in common. 

The Fisher information matrix, given by Equation~\ref{eq:crlb}, is proportional to the probability of observing data $\vec{y_i}$ of stream $i$, given true parameters of the Galactic model $\vec{x}$, that is $C_{x,i}^{-1}\propto \ln(P(\vec{y_i}|\vec{x}))$.
Since the probability for an ensemble of independent experiments is the product of individual likelihoods, it follows that the Fisher information matrix for multiple streams is simply the sum of Fisher matrices for individual streams: $C_x^{-1} = \sum_{i} C_{x,i}^{-1}$.
Note, however, that prior $V_x^{-1}$ should only be included once in the sum.
As in the case of individual streams, the Cram\' er--Rao bounds for multiple streams are given by the inverse of the Fisher matrix, $C_x$.

\begin{figure*}
\begin{center}
\includegraphics[width=\textwidth]{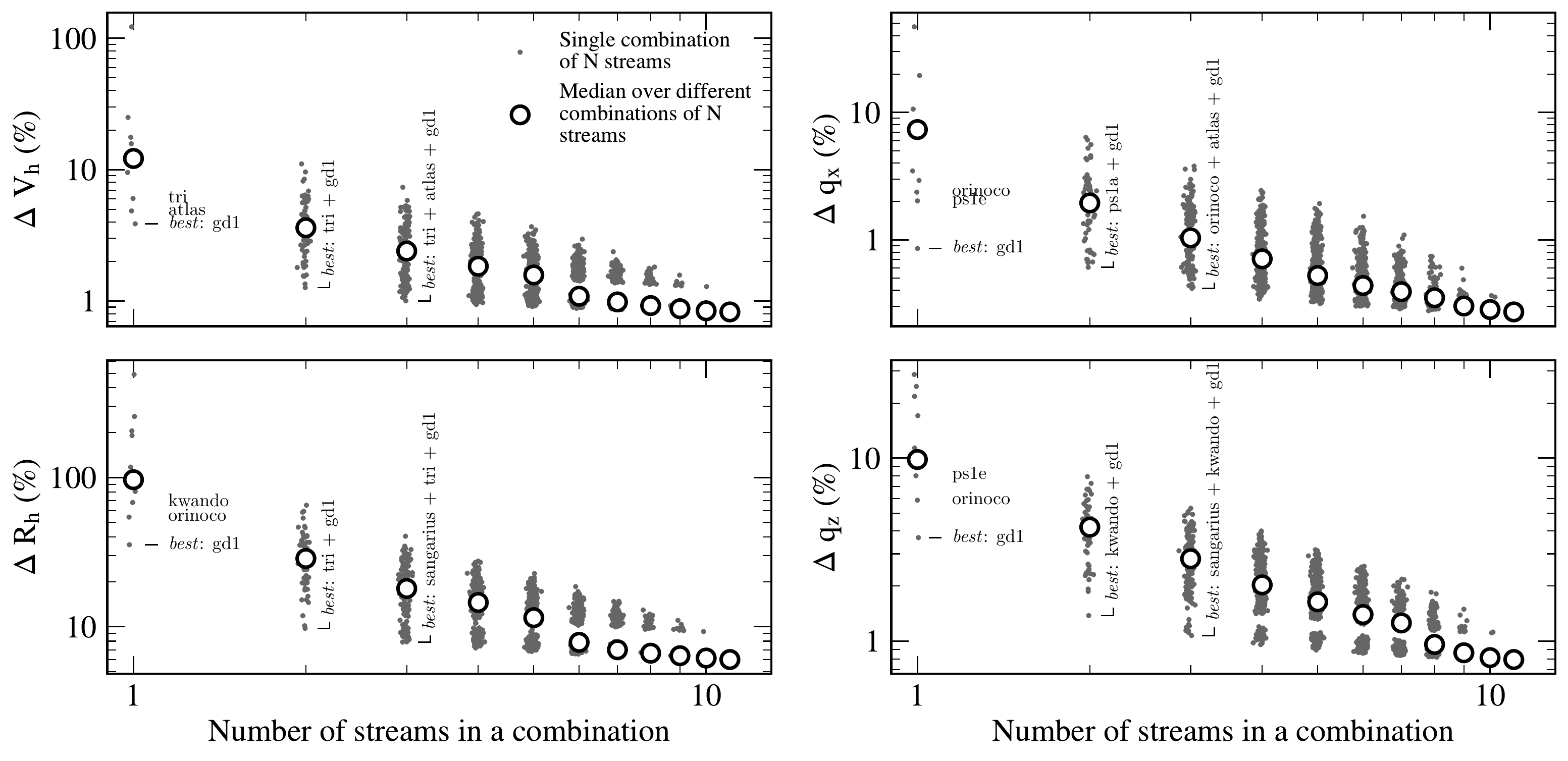}
\caption{Constraints on halo parameters improve when multiple streams are combined.
Each panel shows Cram\'er--Rao bounds on a halo parameter, as a function of the number of streams used to derive the bounds.
Gray points show constraints from a single combination of streams, while white points are the median across all possible combinations for a given number of streams.
Combining any ten streams would provide percent-level precision in all parameters, except the scale radius, which is constrained at a few percent level.
We label the top three streams for each halo parameter, as well as the best-performing pairs and triples.
For the halo scale velocity, the best pair and triple are a combination of streams that top the individual constraints.
However, the information in halo shape parameters is combined in less trivial ways, so that the tightest pair and triple constraints include contributions from streams which are not strongly constraining individually.
}
\label{fig:nstream_summary}
\end{center}
\end{figure*}

There are 11 streams in our sample, and we quantified constraints on the dark matter halo in all possible combinations.
These are summarized in Figure~\ref{fig:nstream_summary}, where every panel shows constraints from a combination of streams on a single parameter of the dark matter halo as a function of the number of streams in a combination (gray points).
There are $\binom{N}{11}$ unique combinations with a total of $N$ streams in a combination, so we highlight the median constraint at a given $N$ with larger white points.
Unsurprisingly, different streams provide somewhat different constraints on the gravitational potential, so joint stream constraints are better than individual ones.
For example, the median \CRLB\ from a triplet of streams is better than the best \CRLB\ from an individual stream in all halo parameters.
Going even further and measuring the gravitational potential with ten streams simultaneously results in the median precision of a few percent for the scale radius (bottom left), while the median precision for halo scale velocity (top left panel), $x/y$ axis ratio (top right) and $z/y$ axis ratio (bottom right) is better than a percent.
Attaining such a precision in the analysis of the Galactic dark matter halo is within reach, and will truly usher in the era of \emph{precision} near-field cosmology.

In addition to studying median constraints for a combination of streams, we also look into which streams are individually performing the best, and also which are a part of the best pair and the best triplet for different halo parameters.
The best three individual streams on each panel are labeled to the right of their respective gray data points, while the best pair and triplet are noted vertically to the right of pair and triplet constraints, respectively.
Scale velocity is best constrained by a combination of streams that constrain it well individually, so members of both the best pair and the best triplet are individually in the top three streams for scale velocity precision.
However, for the rest of the parameters, which determine the halo's shape, it is often the case that at least one of the streams from the most constraining pair or triplet is not individually among the top three most constraining streams for that particular parameter.
In recovering the halo shape, streams complement each other in a non-trivial manner, so the joint constraint is truly more than the sum of its parts.

\section{Applications}
\label{sec:applications}

\subsection{Forecasting and observation planning}
\label{sec:forecast}
\begin{figure*}
\begin{center}
\includegraphics[width=\textwidth]{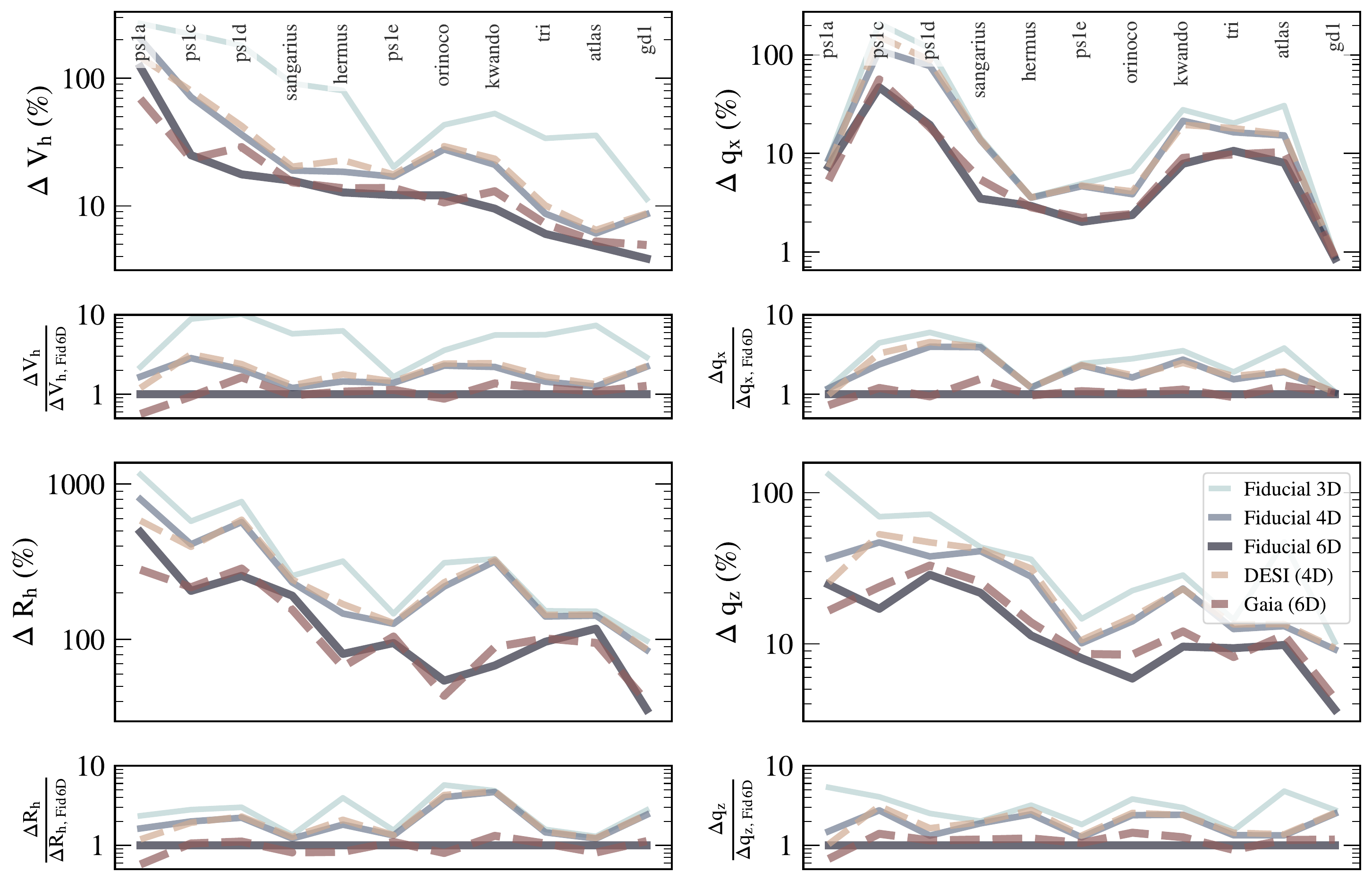}
\caption{Cram\'er--Rao bounds are valuable for planning observations, and in this figure we explore how the precision in recovered potential parameters (y axis) changes for different observing setups (different lines), for all the streams in our sample.
Each panel shows constraints on a single halo parameter (counterclockwise from top left: halo scale velocity, scale radius, $z$-axis and $x$-axis flattening) from different streams, placed along the x-axis, and labeled in the top panels.
In general, constraints get better when the 3D positions are complemented with radial velocities (4D) and proper motions (6D), however, the magnitude of the improvement varies among the streams, as does the relative importance of having 4D versus 6D data.
Observing modes with the same dimensionality of data, but different observational uncertainties (solid and dashed lines of same width), produce similar constraints on halo parameters, hinting that the measurement uncertainty is of secondary importance. 
}
\label{fig:obsmodes}
\end{center}
\end{figure*}

We have so far examined what aspects of the gravitational potential stellar streams constrain if high-quality observational data were available.
Attaining these data would require extensive observational dedication to stellar streams, beyond the level expected in the near term.
However, the \emph{Gaia} mission and a number of spectroscopic projects are slated to soon map a significant fraction of the Milky Way at a lower resolution.
Though in general less precise than the fiducial uncertainties we used so far, these surveys will have the scope required to provide at least one velocity component for most of the known streams in the next few years.
In this section we analyze how constraints on the gravitational potential depend on the availability and quality of observational data along stellar streams. 

The three main scenarios for data availability we considered are: fiducial, DESI-like and Gaia-like, which differ both in the dimensionality of the observational data as well as its precision (for more details, see Section~\S\,\ref{sec:datasets} and Table~\ref{t:datasets}).
In Figure~\ref{fig:obsmodes} we compare Cram\'er--Rao bounds on parameters of the Galactic dark matter halo arising from streams in our sample assuming these kinds of data are available.
Panels in the top and third rows show constraints on a single parameter, with different streams arranged along the x-axis, and ordered by the precision they reach in halo scale velocity, $V_h$, in the fiducial 6D case.
Panels in the second and bottom row show constraints on a single parameter, normalized to the constraints in the fiducial 6D case.
As shown before in Figure~\ref{fig:crb2d_comparison}, different streams are differently sensitive to different parameters of the dark matter halo, and have different requirements on the data quality.

We separately discuss the influence of data dimensionality and data precision on recovery of the halo properties.
\CRLB\ under fiducial observational uncertainties are shown with solid lines in Figure~\ref{fig:obsmodes}, with the thinnest and lightest lines mapping constraints from 3D data, medium from 4D, and thickest and darkest from 6D data.
As expected, higher dimensional data sets provide more precise constraints, with 6D data being on average a factor of 2 better than 4D data, and a factor of 5 better than 3D data.
Halo scale velocity is most susceptible to improvement with high-dimensional data -- switching from 3D to 6D data can improve scale velocity bounds by an order of magnitude.

On the other hand, data precision has only a secondary impact on the recovery of the dark matter halo.
In Figure~\ref{fig:obsmodes} we show the \CRLB\ assuming uncertainties from a DESI-like data set with medium dashed lines, and from a Gaia-like data with thick, dark dashed lines.
DESI-like observations are equivalent to the fiducial 4D observing mode, but assume a factor of 2 larger uncertainties in radial velocities.
This uncertainty, however, has little impact on the recovered \CRLB, which are on average 7\,\% larger than in the fiducial 4D for the halo scale velocity, and only 1\,\% larger for $x/y$ axis ratio of the halo.
Gaia-like observations inherit uncertainties of the fiducial 6D observing mode, but assume distances are known an order of magnitude more precisely, while the proper motions are a factor of 2 worse than in the fiducial case.
Again, these differences in data precision have a modest impact the resulting \CRLB: Gaia-like data is on average slightly less constraining than the fiducial case, but only marginally so.
The largest difference of 20\,\% between the Gaia-like and fiducial 6D data is in halo $z$ axis flattening, whereas the smallest is 3\,\% for $x$ axis flattening.
The dispersion between these two observing mode is 20\,\% -- larger than the median difference, and indeed, Figure~\ref{fig:obsmodes} shows that some streams are more informative with Gaia-like data, while others provide better constraints on the halo with fiducial 6D data.

To summarize, additional dimensions in stream observations improve constraints on the dark matter halo by a factor of few, while the increase in data precision only improves the precision in model parameters by a few percent.
This result, combined with the improvements expected from combining multiple streams (see Section~\S\,\ref{sec:res_joint}) suggests that lower-resolution, large-scale surveys which can target many streams would produce an optimal data set to constrain the Galactic potential with stellar streams.

\subsection{Physical interpretation of stream constraints}
\label{sec:interpretation}
In this work, we analyzed what stellar streams tell us about the Galaxy they orbit, and to do so, we described the Galaxy with a parametric gravitational potential.
This choice made it easy to create models of streams, and study how the distribution of the debris reacts to changes in the properties of the Galaxy.
However, the gravitational potential is a theoretical construct, sourced by the distribution of matter in the Galaxy.
In this section, we relate constraints which streams put on the parameters of the gravitational potential to more physical properties of the Galaxy.

In general, Cram\'er--Rao bounds on a set of variables $\vec{q}$ can be propagated to a new set of variables $\vec{p}(\vec{q})$ using \citep{albrecht2009}:
\begin{equation}
C_p = \left(\frac{d\vec{p}}{d\vec{q}}\right)^{T} C_q \left(\frac{d\vec{p}}{d\vec{q}}\right)
\label{eq:propagation}
\end{equation}
Since the matter density and gravitational potential are related through Poisson's equation, $4\pi\rho = \nabla^2\Phi$, it is possible to translate constraints on the potential field to constraints on the density field directly.
However, calculating the derivatives in Equation~\ref{eq:propagation} in this case would require integrating the Poisson's equation, which would be an involved calculation for a Galactic potential that can in principle be triaxial.
On the other hand, radial acceleration is also a physical quantity, and conveniently, it is a simple derivative of the potential, $a_r = \partial\Phi / \partial r$.
In a spherically symmetric potential, radial acceleration, $a_r(r)$, directly correlates with the enclosed mass, $M(<r)$, as $a_r = G M(<r)/r^2$.
Hence, to better understand physical aspects of the Galaxy that streams constrain, we propagate Cram\'er--Rao bounds on parameters of the gravitational potential to bounds on the radial acceleration at different distances from the Galactic center, with the expectation that bounds on the radial acceleration can be interpreted as bounds on the enclosed mass.

\begin{figure*}
\begin{center}
\includegraphics[width=\textwidth]{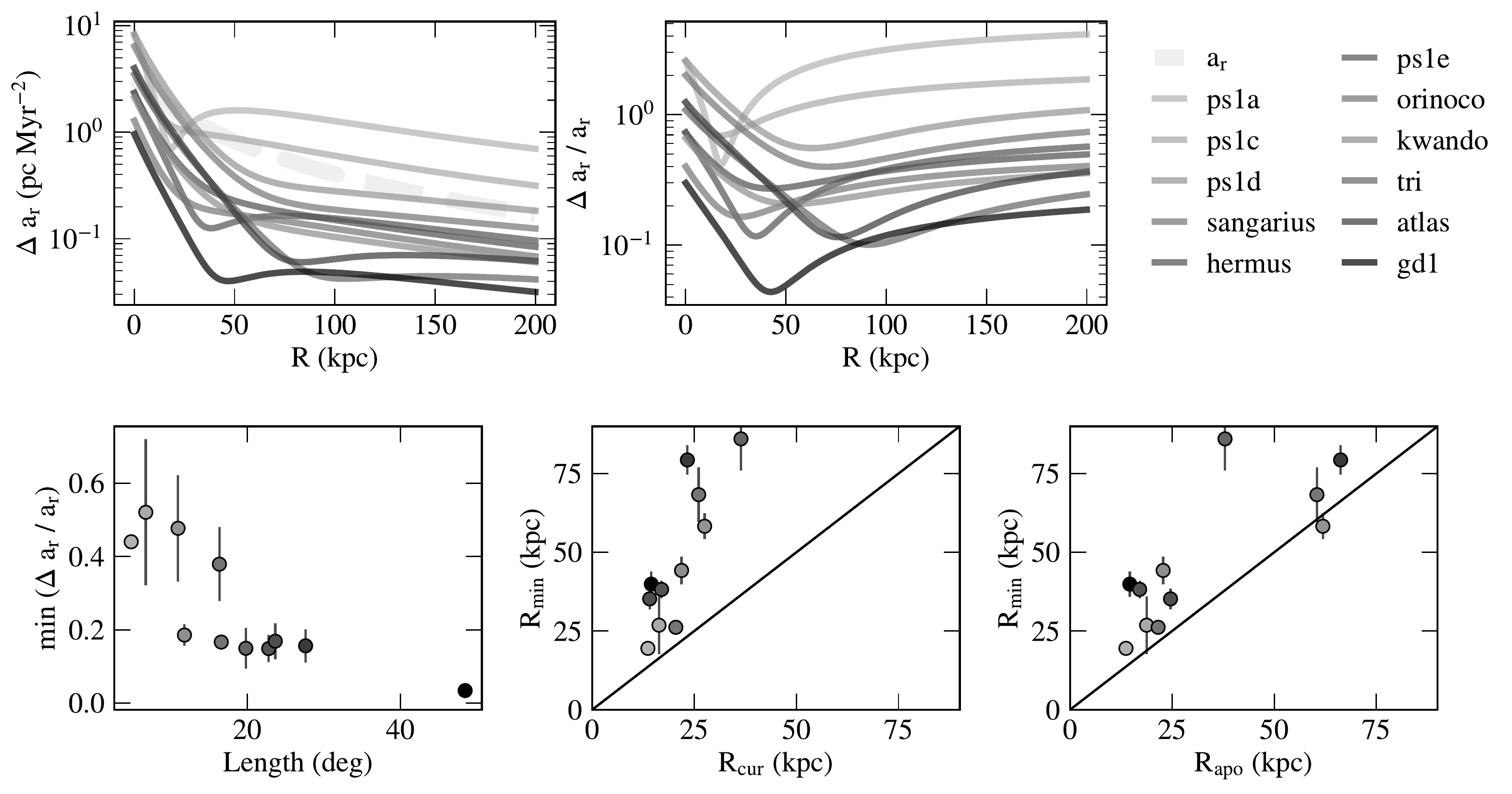}
\caption{Constraints on the radial acceleration by fiducial 6D phase-space data on stellar streams.
Precision in total and fractional radial acceleration as a function of distance from the Galactic center (top left and right, respectively), with the radial acceleration profile shown in dashed gray on the left.
Streams vary in their ability to constrain radial acceleration, but all have a single Galactocentric distance where the fractional constraint is maximized.
Long streams provide tighter constraints (bottom left), and are colored darker in all panels.
Location of the best constraint in the simple model of the Galaxy seems to correlate non-linearly with the present-day position (bottom center) and linearly with the orbital apocenter (bottom right), but see Figure~\ref{fig:ar_all} for constraints on the radial acceleration in more flexible models.
}
\label{fig:ar}
\end{center}
\end{figure*}

The top left panel of Figure~\ref{fig:ar} shows constraints on the radial acceleration as a function of distance from the galactic center along a random line of sight using cold streams observed in the fiducial 6D case (solid lines).
Quantitatively, different streams provide constraints that span an order of magnitude at all radii, consistent with the spread in the \CRLB\ on model parameters (see Section~\ref{sec:forecast}), but in general, stream constraints are qualitatively similar.
For example, the recovery of accelerations is expected to be very uncertain close to the galactic center using any stream, with the typical uncertainty on the order of the acceleration itself (shown as a dashed light-gray line).
This region of the galaxy is baryon-dominated, and we have already found stellar streams to be fairly agnostic on the distribution of baryons (see Section~\S\,\ref{sec:res_ind}), likely due to perceived difficulty in recovering properties of the gravitational potential inside the streams' pericenter, which then drives this uncertainty in accelerations.
At large radii, constraints on radial acceleration improve for all streams, with the \CRLB\ for some objects featuring a local minimum before continuing to decrease.
The overall acceleration is also decreasing, so we show \CRLB\ on the radial acceleration relative to the acceleration in the top right.
The relative uncertainty on the acceleration is large at small radii, it then decreases to a local minimum and asymptotically increases at large radii.
The exact profile varies between different streams, but the presence of a local minimum is universal, indicating that the information streams provide about the gravitational potential is radially confined.

In the bottom panels of Figure~\ref{fig:ar} we explore in more detail the best constraints on the acceleration achievable using streams in our sample.
Since the galaxy model is not spherically symmetric, we calculated constraints on the radial acceleration along 50 random lines of sight and marked the median with circles, while the errorbars indicate 16th and 84th percentile in all three panels of the second row.
Bottom left panel presents the best relative \CRLB\ for a given stream as a function of the stream length.
Similarly to constraints on the parameters of the galactic potential, the best constraints on the acceleration are generally better for long streams (shown in darker colors), although there is some scatter between different streams, and also for a given stream among different lines of sight.
We next consider where these best measurements of the acceleration are obtained, parameterized with the location of the local minimum in the fractional constraints on the radial acceleration, $R_{min}$.
In the middle bottom panel we show $R_{min}$ as a function of current radial distance of the stream from the Galactic center, $R_{cur}$, and on the bottom right as a function of the progenitor's apocenter, $R_{apo}$.
In both cases, positions of the best stream constraints are scattered off the one-to-one line (solid black), but they in general seem to correlate with both quantities.
Different functional forms relate $R_{min}$ to $R_{cur}$ and $R_{apo}$ (non-linear and approximately linear, respectively), so it is unclear which, or if either, of these relations is intrinsic.

Understanding how the location of a stream's best constraint is set would help in interpreting what the streams are actually measuring.
For example, attaining the best measurement of the radial acceleration at the apocenter would imply that streams at any point in time contain information on their whole orbital history, constraining the total mass enclosed within their apocentra.
Alternatively, if the best measurement is at a stream's current position, that means that streams are foremost a local measure of the enclosed mass.
Uncovering the true relation in our fiducial, parametric model of the Galaxy is hindered by the structure that such a model imposes on all spatially-resolved constraints.
As a final application of our framework to measure information content in streams, in the next section we relax our assumptions about the gravitational potential and distinguish between these radically different interpretations using a more flexible potential model.

\subsection{More flexible potential models}
\label{sec:bfe}
One of the main findings in this work is that combining stellar streams provides an extremely precise handle on properties of the dark matter halo, with individual parameters of a triaxial NFW model being jointly constrained to $\lesssim1\%$ (Section~\S\,\ref{sec:res_joint}).
At this level of precision, NFW is no longer an adequate representation of the underlying potential field \citep[e.g.,][]{bonaca2014}, which additionally motivates the expansion to more complex and more flexible models.
In this section, we explore stream constraints on the radial acceleration when the basic model of the Galaxy is allowed to be perturbed.

We introduce more flexibility into the gravitational field that streams orbit by considering a potential model that is a combination of the Galactic potential and a perturber: $\Phi = \Phi_{G} + \Phi_{p}$. 
For simplicity, we assume that the perturbation is caused by external multipoles, whose contributions to the potential field are given by:
\begin{equation}
\Phi_{p}^{l_{max}}(r) = \sum_{l=1}^{l=l_{max}}\sum_{m=-l}^{m=l} a_{lm} Y_{lm}(\theta,\phi) r^l
\label{eq:multipoles}
\end{equation}
where $r$ is the distance from the center of the Galaxy, $Y_{lm}$ are spherical Bessel functions, $l_{max}$ determines the highest order perturbation, and coefficients $a_{lm}$ are the additional parameters of this model.
We choose vanishing coefficients, $a_{lm}=0$, such that this flexible potential is equivalent to the fiducial Galaxy we analyzed so far, but it has the additional model freedom.
In principle, a multipole decomposition of a high order ($l\approx 10$) can reproduce the structure of dark matter halos expected in the $\Lambda$CDM paradigm even on small scales \citep{lowing2011}.
We limit the scope of this section to merely illustrate the effects of loosening the potential structure by including perturbations up to octupole order, $l_{max}=3$.
A perturbation of order $l$ adds $2\,l+1$ additional parameters to the model, so our extended model has at most 15 new parameters in addition to the 15 from the basic study (listed in Table~\ref{t:model}).
In what follows, we explore how this additional freedom propagates into recovery of the underlying acceleration field using Milky Way-like stellar streams.

\begin{figure*}
\begin{center}
\includegraphics[width=\textwidth]{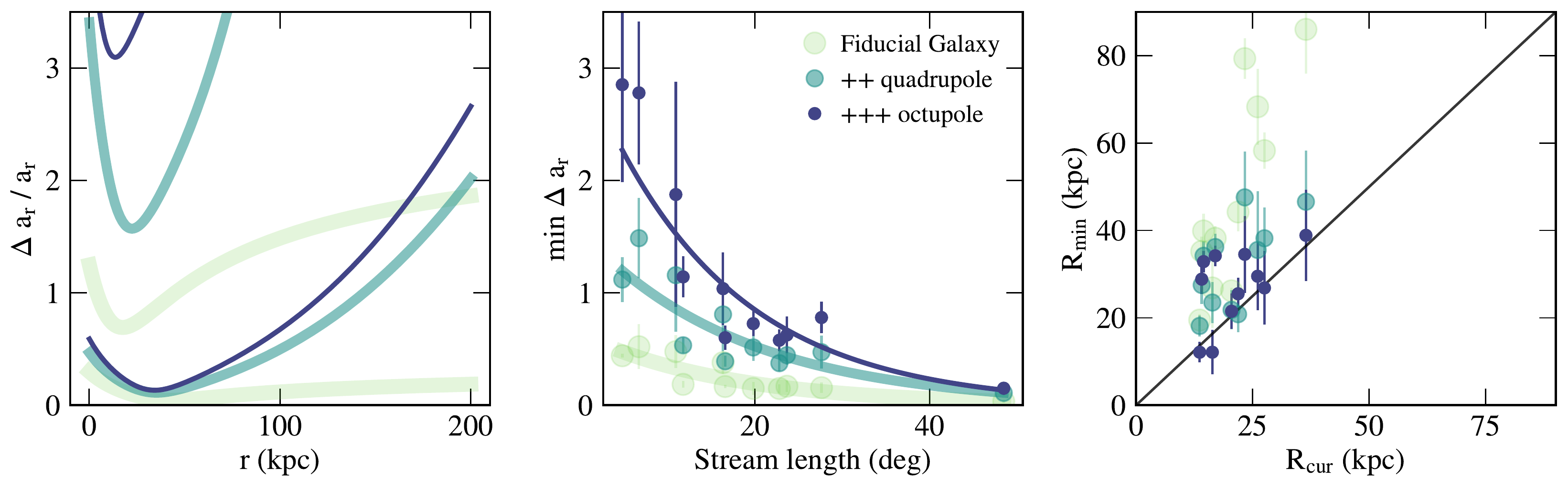}
\caption{
Fractional constraints on the radial acceleration from 6D data on stellar streams in models of increasing complexity: light green for the basic model of the Galaxy, dark green for additional dipole and quadrupole freedom, and purple for freedom up to the octupole term.
Radial profiles of the acceleration constraints are shown for a couple of streams on the left, dependence of the best measured acceleration on the stream length is in the middle, and locations of best constraints as a function of current stream position are on the right.
In more flexible models, observations of streams provide less information on the radial acceleration, especially if a stream is short.
However, modeling streams with sufficient freedom also enables them to measure the local acceleration at their current position.
In this regime, combining constraints from many different streams becomes computationally tractable.
}
\label{fig:ar_all}
\end{center}
\end{figure*}

Figure~\ref{fig:ar_all} shows constraints on the radial acceleration as the potential model gets progressively more complicated; light green symbols indicate our basic model of the Galaxy, dark green is for the basic model with a dipole and a quadrupole freedom, while purple is for a model that includes up to the octupole term.
Left panel shows the relative precision in radial acceleration as a function of distance from the galactic center for an arbitrary sightline, and to reduce clutter, we only present constraints from two streams.
Unsurprisingly, the same amount of data constrains models with more parameters less well overall, so the acceleration profiles of more flexible models have larger \CRLB\ at all radii. 
The precision of the best measured relative acceleration increases exponentially with stream length (shown for all streams in the middle panel of Figure~\ref{fig:ar_all}), and the scale length is $l_0\approx17^\circ$ in all models.
For streams of the same length, the best acceleration measurement in a model with added dipole and quadrupole freedom is a factor of two worse than in the basic model, while additional octupole component increases this uncertainty by a factor of five relative to the basic model.

While the uncertainty in recovery of all parameters increases with model flexibility, constraints on the radial acceleration also become more localized in more flexible models.
Radial profiles of precision in the radial acceleration (left panel of Figure~\ref{fig:ar_all}) have deeper minima in progressively more flexible models, which indicates that each stream intrinsically measures only a specific region of the Galaxy.
Bounds on the radial acceleration in the basic model indicated that streams are the most constraining within their apocentra (see bottom right of Figure~\ref{fig:ar}), however, this seeming relation dissolves in less constrictive models of the Galaxy.
Instead, we find that progressively more complex models move the best measured location closer to the current location (Figure~\ref{fig:ar_all}, right).
In our most flexible model, the best constraints for all but a few streams are attained within $1\,\sigma$ of their current position.

Augmenting the basic model of the Galaxy with external multipoles is a naive first step towards a spatially flexible and realistic model for the Milky Way, but it already afforded the physical insight that stellar streams are most sensitive to its present environs.
The finding that streams constrain the mass enclosed within their current position, with a precision determined by their length, has far-reaching implications for observational studies as both the position and length are directly observable.
Aimed with this knowledge, we can tailor observational experiments and prioritize telescope resources to study the most informative streams.

\section{Discussion}
\label{sec:discussion}
This paper performs an information analysis of cold stellar streams, and quantifies how streams on different orbits constrain global parameters of the dark matter halo.
In \S\,\ref{sec:dis_origin} we reflect on results relating stream properties to aspects of the gravitational potential they constrain.
In this first analysis of its kind, we made a number of simplifying assumptions, which we critically revisit in \S\,\ref{sec:dis_caveats}.
We briefly discuss possible extensions to the basic framework in \S\,\ref{sec:dis_extensions} and close by considering the implications for future studies of streams in \S\,\ref{sec:dis_applications}.

\subsection{Origin of the information in stellar streams}
\label{sec:dis_origin}
We started this project with two questions in mind:
First, what kind of data do we need in order to map the dark matter halo using stellar streams?
Secondly, what aspects of the halo do these streams constrain?
Answering these questions by calculating the Fisher information for eleven Milky Way-like streams opened up an avenue for addressing a myriad of new questions.
Many of these can be explored using the Fisher matrices from this work, which are available at \url{https://github.com/abonaca/stream_information}.\ 
In this section we reflect on our findings so far, and indicate directions we find especially interesting for future study.

In response to our practically minded consideration of data required to fit streams, we find that some kinematic information is essential for improving upon the existing measurements of the Milky Way's gravitational potential.
The degree to which kinematics improve potential constraints depends on the stream, but most streams display the most marked improvement when a single component of the velocity vector is available, while having the full vector, for example proper motions in addition to radial velocities, provides a minor refinement.
Similarly, an increase in the precision of measured velocities barely has an effect on potential constraints.
These results indicate that in rigid models for the gravitational potential, most of the model degeneracies can be lifted with just a single projection of the 3D bulk-velocity vector.
However, we expect that precise data and full 3D velocity information will significantly improve constraints in more flexible potential models  -- a hypothesis to be tested in the future.

In response to our second question, we have shown throughout this paper that what we learn about the Galaxy depends on what streams we analyze.
In general, longer streams are more informative about all aspects of the gravitational potential, and are prime targets for more detailed follow-up.
Additionally, we find that streams on eccentric orbits are more informative about the shape of the dark matter halo, while streams orbiting in the inner part of the Galaxy can provide tighter constraints on properties of its stellar disk.
This correlation between the volume that streams orbit and the properties of the potential they constrain is further affirmed in constraints on the radial acceleration.
In our fiducial, parametric model of the Galaxy, the radial acceleration is best measured at distances similar to where the stream resides currently, but in more flexible potentials the best constraints truly hone in on the current location of the stream.
In general, we expect more freedom in the model to allow for perturbations away from the stream, as long as they cancel out at the location of the data.
With a moderately flexible model, we were able to demonstrate this localization of stream information in the radial direction, whereas future studies with fully flexible potentials will test whether stream constraints are local in all three dimensions.

\subsection{Caveats regarding the current framework}
\label{sec:dis_caveats}
The Fisher framework for assessing the information contained in a given set of observables has the intrinsic limitation of being tied to a model.
In our case, this means that we are not measuring the information in the observed streams, but rather in toy streams that are comparable to the streams in the Milky Way.
Given that there are no kinematic measurements for most of these streams, our choice for the direction of motion in the toy stream is somewhat arbitrary.
The information content in streams depends on properties of the progenitor to some extent, so our results serve to illustrate how streams constrain the potential in general, rather than to make detailed predictions for these exact streams.
Furthermore, by testing how sensitive stream observables are to changes around the fiducial potential we only measured the precision with which streams would constrain parameters of the gravitational potential.
These predictions are robust to the choice of a fiducial model; for example, in prior applications to the cosmic microwave background, the precision in recovered parameters of the $\Lambda$CDM model matched the predicted Cram\'er--Rao bounds, even though they were calculated in a model with different fiducial parameters \citep{bond1997,spergel2003}.
However, without the full mapping of the likelihood surface, we have no handle on biases that might arise in potential recovery using streams.

The Fisher analysis is performed in the context of a model because it ultimately relies on calculating derivatives of the data with respect to the parameters of the model, and these derivatives are evaluated at the fiducial values of the model parameters.
In some problems, the derivatives are known exactly, but the transformation from the parameters of the gravitational potential and position of the stream's progenitor to the properties of the stream is highly non-linear, and there are no analytic derivatives that connect these two sets.
For simplicity, we employed numerical derivatives, and chose steps of an appropriate size to ensure their convergence for all the model parameters (see \S\,\ref{sec:derivatives}).
This conservative choice guarantees that the \CRLB\ are not overly optimistic due to noisy derivatives.
Still, the fact that our derivatives are numerical is the main technical limitation of this work, and in the next section we outline how to improve upon this.

Another caveat of this work is that its stream data model is only a naive reflection of the more complicated reality. 
In particular, we assume the observations are normally distributed and uniformly sample the stream, the uncertainties are homoscedastic, and that the stream members are identified with perfect confidence. 
In reality, measurements are typically made in the densest parts of a stream, their uncertainties vary (e.g., brighter stars are measured better), and they might not be Gaussian in all of the observables (e.g., if distance information is coming from parallaxes, then uncertainties in distance are not Gaussian, even if the parallaxes themselves are normally distributed).
Additionally, if stream kinematics overlap with the Milky Way field population, it is impossible to reliably identify stream members without further chemical information, so instead we can only label stars with a membership probability.
However, we expect the average stream track to be measured well in a six-dimensional space for all of the nearby streams in the \emph{Gaia} era, so even though this basic model fails to capture complexities of observational data, it is on average credible, and can easily be relaxed and tailored for forecasting specific data sets.

Finally, we note that the analytic gravitational potential we used can hardly account for the dynamics of the Milky Way over extended periods of time.
In the early universe, we expect the Galaxy to have evolved through constant merging which triggered migrations within the already formed stellar populations \citep[e.g.,][]{elbadry2016,bonaca2017}, while at the present the motions of stars are affected both by the Sagittarius dwarf galaxy \citep{laporte2017} and by the Magellanic system \citep[e.g.,][]{gomez2015,laporte2018a}.
A simple decomposition of the Galaxy into a bulge, disk and halo component has been the only tractable method of inferring the gravitational potential \citep[e.g.,][]{lm10,koposov2010,bovy2016}, but due to the increase in both precision and volume of potential tracers, it is fast reaching the limit of its applicability (see Section~\S\,\ref{sec:res_joint}).
Inspired by the expansion we have begun in Section~\S\,\ref{sec:bfe}, in the following section we suggest a way forward using a more flexible representation of the underlying gravitational field.

\subsection{Extensions of the basic forecasting framework}
\label{sec:dis_extensions}
In the previous section we identified the use of numerical derivatives and overly idealized forms of the gravitational potential as prime targets for improving our inference of the information in stellar streams.
The numerical derivative of stream observable $y_i$ with respect to the model parameter $x_i$ measures the response of the observable $\Delta y_i$ to the change in parameter $\Delta x_i$, and is calculated by differencing the observable $y_i$ between two stream models separated by $\Delta x_i$ at a fixed position along the stream.
In principle, one can imagine using the orbit integrator to directly propagate the change in parameter $x_i$ into observable $y_i$, instead of creating separate stream models to evaluate this difference.
Indeed, such functionality, called auto-differentiation, is common in machine-learning codes \citep[e.g., \package{tensorflow},][]{tensorflow}, and also exists in some N-body codes \citep[e.g., \package{FASTPM},][]{feng2016}.
In this work we opted for evaluating numerical derivatives because our legacy code does not support auto-differentiation, however, implementing it tops our list of future extensions.
Principal derivatives of stream observables with respect to model parameters will not only make for more robust \CRLB, but could also aid in fitting streams, as higher-order parameter-estimation methods, such as Hamiltonian Monte Carlo, rely on evaluating both the likelihood and the likelihood derivative.

To fully describe a population of streams, we need a realistic model of the Galaxy.
The latest generation of hydrodynamical simulations have produced models that well reproduce a multitude of features observed in galaxies, both as individual objects \citep[e.g.,][]{wetzel2016} and as a population \citep[e.g.,][]{pillepich2018,nelson2018}.
Part of their success is in the achieved high resolution, so at the present a Milky Way-like galaxy is modeled by up to 140 million particles \citep{wetzel2016}.
Ultimately, we would like to have a description of a galaxy that is representative of these models, but at a fraction of the numerical cost.
In the classic study, \citet{ho} developed a set of basis functions for density and gravitational potential that can reproduce complex morphology of galaxies with a small number of terms \citep[e.g.,][]{lowing2011, lilley2018a, lilley2018b}.
These expansions reproduce the force field of an N-body simulation with a precision of a few percent, so representing the gravitational potential in our model with an expansion of basis functions is tempting.
However, a truly realistic solution needs to accurately capture not only the current structure of the galaxy, but also its evolution in time.
Even though the Milky Way has had a relatively quiet recent merger history, it is currently undergoing a major merger with the Large Magellanic Cloud \citep[e.g.,][]{besla2007, penarrubia2016}, which has been a source of gravitational perturbation for at least a billion years -- sufficiently long to affect stellar streams.
Specifically, \citet{lm10} showed that the only static, ellipsoidal halo that reproduces both the positions and radial velocities along the Sagittarius stream is triaxial.
This model correctly predicted proper motions along the stream \citep{sohn2015}, so it appears to be describing the effective potential well, even though it is cosmologically improbable \citep{debattista2013}.
On the other hand, modeling Sagittarius in a combined system of the Milky Way and the LMC relaxes the requirement for the dark matter halo to be triaxial \citep{vch2013}, signaling that having a model of the potential that is correct on average is no guarantee of recovering the true halo shape.
To ensure that the complexities added to the model are realistic, the basis function expansion should thus be time-dependent, simultaneously describing the interaction between the Milky Way and the LMC, while maintaining a thin and old disk. 
We delegate the development of such a model and its implementation for mapping the dark matter in the Galaxy to a future study.

\subsection{Implications for stream studies in the Milky Way and beyond}
\label{sec:dis_applications}
We studied the information content in stellar streams with the dual intent of learning what they are telling us about the gravitational potential intrinsically, and what we can measure about the gravitational potential using a given set of data.
Such forecasting, while standard in other areas of astronomy \citep[e.g.,][]{bond1997,tegmark2000}, is rather novel in near-field cosmology.
The stream studies in particular were until now limited by the scarcity of sources, so individual objects were dedicated the individual follow-up and subsequent modeling.
However, advances in discovery techniques applied to the legacy data \citep{grillmair2014,grillmair2017a}, as well as the emergence of deeper data \citep{shipp2018}, have increased the count of known stellar streams to a few dozen \citep{grillmair2016}.
The majority of these new detections have no associated kinematics, a trend likely to be extended in the LSST era, when the number of objects to observe spectroscopically, both in the Milky Way and other galaxies in the local universe, will be prohibitively large.
In this new regime, it will be essential to identify the most efficient targets among the streams, and also within the stream itself, to optimally use the shared resources -- a task ideally suited for the framework developed in this work.

In addition to resulting in a tool for observation planning, measuring the information content in streams gave us pointers on ways to optimally extract it.
\begin{itemize}
\item{First, we should infer the gravitational potential using a population of streams, because as a group they outperform any individual stream by more than an order of magnitude (see Figure~\ref{fig:nstream_summary}).}
\item{Second, streams are measuring local properties (see Section~\S\,\ref{sec:interpretation}), so we should describe the gravitational potential with a very flexible model that can adequately capture information provided by streams without imposing artificial structure.}
\item{Third, if using sufficiently flexible model, we should be able to analyze each stream individually and measure the acceleration field at their respective locations.
The inference on the global potential field can then be done as a separate step of interpolating between these individual measurements.}
\end{itemize}
The approach we outlined capitalizes on the hierarchical structure of the problem, which makes it computationally more efficient than simultaneously fitting all of the streams, and at the same time it not only enables, but requires a realistic representation of the Milky Way.

\acknowledgements{
This work was improved upon following the thoughtful suggestions of the Conroy group at Harvard, the Finkbeiner group at Harvard, the NYC Stars group convening at the Flatiron Institute, Dan Foreman-Mackey, Kathryn Johnston, Nikhil Padmanabhan, Adrian Price-Whelan, Hans-Walter Rix, Hy Trac and Dan Weisz, and it is a pleasure to thank them.
AB acknowledges generous support from the Institute for Theory and Computation at Harvard University; and the Max-Planck-Institut f\"ur Astronomie and the Flatiron Institute Center for Computational Astrophysics, where parts of this work were developed.
DWH was partially supported by the National Science Foundation (grant AST-1517237).
All code used in this work and all results are available at \url{https://github.com/abonaca/stream_information}.
}

\software{
    \package{Astropy} \citep{astropy},
    \package{gala} \citep{gala},
    \package{h5py} (\url{https://www.h5py.org/}),
    \package{IPython} \citep{ipython},
    \package{matplotlib} \citep{mpl},
    \package{numpy} \citep{numpy},
    \package{scipy} \citep{scipy}
}

\bibliographystyle{aasjournal}
\bibliography{crlb}

\appendix
\section{Cold tidal streams in a Milky Way-like galaxy}
\label{sec:streams}
To study the information content in stellar streams, we generated a set of mock streams that have similar properties to the known streams in the Milky Way.
We limited our study to eleven streams in the PanSTARRS footprint: ATLAS, GD-1, Hermus, Kwando, Orinoco, PS1A, PS1C, PS1D, PS1E, Sangarius and Triangulum.
These streams range in length from a few to more than $50^\circ$ and in Galactocentric distance from 15 to 35\,kpc, thus representing some of the diversity in the currently known population.

\begin{figure}
\begin{center}
\includegraphics[width=\textwidth]{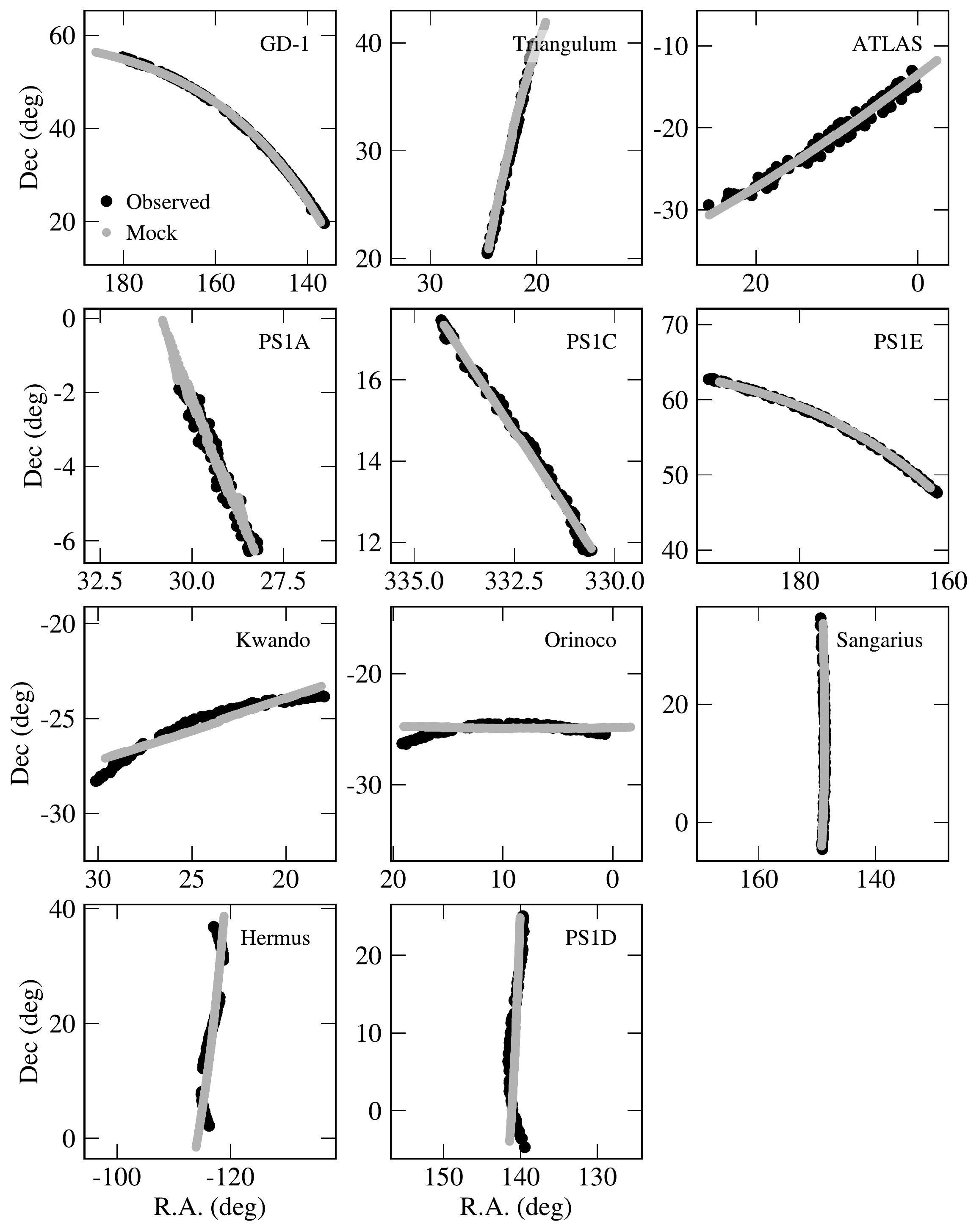}
\caption{On-sky positions of streams analyzed in this work, with one stream per panel, and the stream name in the upper right corner.
In each panel, the observed stream members are shown in black, whereas a best-fitting mock stream is plotted in gray.
All of the mock streams have been generated in the same gravitational potential, and they fit the observations reasonably well.
Streams where the mock deviates from observations likely experienced a different gravitational potential for at least a part of its evolution.
}
\label{fig:gallery}
\end{center}
\end{figure}

Mock streams resembling observed objects, such as the ones used in this work, can be created following these steps:
\begin{itemize}
\item Obtain ridge points along the observed stream.
For some streams, mean tracks have been published, usually in a polynomial form, and we used published tracks for Sangarius from \citet{grillmair2017a}, Kwando and Orinoco from \citet{grillmair2017b}, Hermus from \citet{grillmair2014} and Triangulum from \citet{bonaca2012}.
For other streams, tracks can be obtained by directly analyzing a match-filtered map showing the stream, and we analyzed maps from \citet{bernard2016} to get positions of ATLAS, GD-1, PS1A, PS1C, PS1D, and PS1E.
 
\item Fit a polynomial to the ridge points, and define a stream coordinate system with one axis along the polynomial, and the other along the normal on the polynomial.

\item Query a star catalog and assign a membership probability \citep[we used PanSTARRS1,][]{ps1,schlafly2012,schlafly2011}.
We assume that the total probability can be split in a spatial and color-magnitude component, $p = p_{spatial} \times p_{CMD}$.
The spatial probability follows a normal distribution centered on the stream track (measured as described above), with dispersion matching the reported stream width.
The color-magnitude probability is obtained from a matched filter based on the M13 globular cluster, and placed at the reported distance to the stream.

\item Select most likely members.
We decided to work with a hunderd stars per stream, and replaced the lowest probability members by stars with kinematic measurements, if any were available.
Sky positions of the adopted stream members are shown in black in Figure~\ref{fig:gallery}.

\item Constrain parameters of the mock stream progenitor so that it produces a stream matching the observed distribution of member stars.
We created models of mock streams in a fiducial gravitational potential defined by the first 9 parameters listed in Table~\ref{t:model} and these parameters were kept fixed for all of the mock streams.
The parameters we varied were: 6D position of the progenitor today, initial mass of the progenitor (we assume the progenitor loses mass at a constant rate and disappears at the present), and the age of the stream.
We estimated these parameters following \citet{bonaca2014} and show the adopted mock streams in gray in Figure~\ref{fig:gallery}.
\end{itemize}
In most cases, mock streams track well the positions of their observed counterparts.
For some, however, the fit is not particularly good; for example, we can't seem to reproduce the observed curvature of Kwando and Orinoco in our fiducial model.
This discrepancy indicates that the true gravitational potential of the Milky Way is more complex than the simple model we adopted as our fiducial, and that just the positions of a population of streams can rule out some models \citep[cf.][]{pearson2015}.
This work focused on assessing the precision, and not accuracy, of stream constraints, so it is acceptable that our fiducial potential is only approximately representative of the Milky Way.

\section{Robust matrix inversion}
\label{sec:inversion}
Inverting a Fisher matrix is a core operation in calculating Cram\' er--Rao lower bounds and thus measuring the information content in stellar streams. 
Often, this problem is ill-conditioned, i.e. if the input Fisher matrix $I$ is modified only slightly, the standard numerical algorithms, such as that employed in \texttt{numpy.linalg.inv}, return a very different inverse $I^{-1}$.
We adopt an iterative approach of rescaling the problem to ensure a robust inverse is obtained, and outline it below.

Starting with a matrix $A$, which may have a large condition number, we are looking for a matrix $Q$, such that $Q A = I$.
If we have a guess for $Q$, and the guess is good, then the matrix $QA$ has a low condition number, and can be reliably inverted using the standard algorithm for numerical inversion.
Then it follows:
\begin{equation*}
(QA)^{-1} Q = A^{-1} Q^{-1} Q = A^{-1}
\end{equation*}
We have just obtained a better guess for the inverse of the original matrix $A$, and adopt it as the matrix $Q$ in the next iteration.
This procedure is repeated until $Q$ converges to $A^{-1}$, i.e., $Q A = I$ to machine precision.

In our current implementation, we use the standard, unreliable inverse as the starting guess $Q$. 
Convergence to the true inverse is typically obtained after several ($\lesssim2$) iterations.

\end{document}